\documentclass[12pt]{iopart}

\usepackage{graphicx}

\usepackage{iopams}
\usepackage{amsfonts}

\begin{document}

\title[Dirac particle II]
{Classical Dirac particle II. Interaction with an electromagnetic plane wave}

\author{Juan Barandiaran}
\address{Telecommunication Engineer, Bilbao, Spain}
\ead{barandiaran.juan@gmail.com}
\author{Mart\'{\i}n Rivas}
\address{Theoretical Physics Department, The University of the Basque Country,\\ 
Bilbao, Spain}
\ead{martin.rivas@ehu.eus}

\begin{abstract}
In a previous work we have described the classical structure and analyzed the interaction of the classical Dirac particle with uniform and oscillating electric and magnetic fields. In the present paper we consider the interaction of the Dirac particle at rest with an electromagnetic  plane wave packet. The integration of the dynamical equations is done numerically with a {\it Mathematica} notebook that is available to the reader. Independently of the electric charge of the Dirac particle the interaction produces a linear momentum transfer such that the center of mass of the particle is moving with a component of its CM velocity in the direction of the propagation of the wave. The electromagnetic plane wave transfers energy, linear momentum and angular momentum to the particle. The CM spin is also modified and its change depends on the torque of the external Lorentz force defined at the center of charge position with respect to the center of mass and on the work of the external force along the center of mass trajectory of the particle. We analyze how the different physical parameters of the wave and the spin orientation determine the outcome of the numerical experiment.
\end{abstract}

\hspace{1.2cm}{\small\bf Keywords:} {Spinning electron; Dirac particle; plane wave}

\section{Introduction}
\label{intro}
In this article we analyze the interaction of the classical spinning Dirac particle described in detail in the previous paper \cite{classical1}, under an electromagnetic plane wave. The analysis of its classical structure and the dynamical equations of this particle are described in that paper which we submit to the reader for their description. The complete theory of elementary spinning particles is described in the reference \cite{Rivasbook}. The main feature is that this elementary spinning particle is described by the evolution of a single point ${\bi r}$, the centre of charge of the particle (CC) which satisfies a system of ordinary fourth order differential equations. The particle also has another characteristic point ${\bi q}$, the center of mass (CM) such that the fourth order system of differential equations for the point ${\bi r}$ can be decoupled into a system of ordinary second order differential equations for both points ${\bi q}$ and ${\bi r}$, where the CM satisfies Newton-like dynamical equations in terms of the external electromagnetic force defined at the CC. We shall compare the motion of the Dirac particle with the motion of a spinless point particle under the same electromagnetic interaction, to distinguish how the spin is affected by the interaction. In Section {\bf\ref{Natural}}
we introduce a natural system of units, where for the electron $m=1$, $S=1/2$, $e=1$ and the universal constants $\hbar=1$ and $c=1$. Section {\bf\ref{wave}} describes the circularly polarized electromagnetic plane wave which is going to interact with the Dirac and point particle. The Section {\bf\ref{eqs}} is devoted to the description of the dynamical equations in the presence of an external electromagnetic field and the considerations relative to the different orders of magnitude of the intensity and frequency of the wave and of the physical parameters that characterize the interaction of the electron with the polarized wave. The Section {\bf\ref{Variation}} is a summary of the spin dynamical equations to justify how the spin is changing when the particle is accelerated. Finally, in Sections {\bf\ref{Inter}} and {\bf\ref{point}} we analyze the interaction of the Dirac particle and the spinless point particle, respectively, with an external electromagnetic plane wave. We end the paper in Section {\bf\ref{Conclusions}} with some general conclusions.
\section{Natural units}
\label{Natural}
In the above work \cite{classical1} we defined a natural system of units: the speed of light $c=1$, the natural unit of length $2R_0=\hbar/mc=1$, that represents twice the separation $R_0$ between the CC and the CM of the Dirac particle if $m$ is the mass of the electron. This defines a natural unit of time $\tau_0=2R_0/c$, so that all boundary variables of the Dirac particle are described by dimensionless variables. 
\[
\widetilde{\bi r}={\bi r}/2R_0, \quad \widetilde{\bi u}={\bi u}/c,\quad {\widetilde{\bi q}}={\bi q}/2R_0, \quad \widetilde{\bi v}={\bi v}/c,\quad \widetilde{t}=t/\tau_0.
\]
We shall delete the tildes from now on and we consider that the variables are always written in natural units. From the mechanical point of view we need an extra natural unit of mass. If we assume that Planck's constant $\hbar=1$, then
\[
2R_0=\frac{\hbar}{mc}=1,\quad\Rightarrow\quad m=1{\rm n.u.},
\]
therefore the natural unit of mass is that unit where the mass of the electron is $1$. 
The energy of the particle is expressed in terms of the center of mass velocity as $H(v)=\gamma(v)mc^2$ which in natural units for the electron $H(v)=\gamma(v)$. The linear momentum is expressed as ${\bi p}(v)=\gamma(v){\bi v}$. The value of the spin in the center of mass frame $S=\hbar/2=1/2$ in natural units. The electron has as intrinsic parameters $m=1$ and $S=1/2$ in natural units. The general expressions of both spins for the electron in natural units are
\begin{equation}
{\bi S}=-\gamma(v)({\bi r}-{\bi q})\times{\bi u},
\label{sCC}
\end{equation}
\begin{equation}
{\bi S}_{CM}=-{\gamma(v)}({\bi r}-{\bi q})\times({\bi u}-{\bi v}),
\label{sCM}
\end{equation}
where the variables ${\bi r}$, ${\bi q}$, ${\bi u}$ and ${\bi v}$ are expressed in natural units. In general, the absolute value of both spins will be a function of the center of mass velocity ${\bi v}$, $S(v)$ and $S_{CM}(v)$,
with the values at rest  $S(0)=S_{CM}(0)=1/2$. The dynamics modifies the variables they depend  and the final value will be determined during the dynamical process. Part of the analysis is to determine how the CM spin is modified by the interaction as we shall discuss in Section {\bf\ref{Variation}}.

From the electromagnetic point of view we need a natural unit for the electric charge. In this case the electric charge of the electron is defined through the fine structure constant by redefining the permitivity of the vacuum $\epsilon_0$.
\[
\alpha=\frac{e^2}{4\pi\epsilon_0 c\hbar}=0.007297, \quad\rightarrow\quad e=1\;{\rm n.u.}\quad\Rightarrow \quad \frac{1}{4\pi\epsilon_0}=\alpha.
\]
What we have is to translate the international system of units to this natural system of units.
The relationship for the fundamental units of mass [M], length [L], time [T] and electric charge [Q] is:
\[
1\; {\rm n.u.} [M]=m_e=9.109534 \cdot10^{-31}{\rm Kg},\quad\rightarrow 1\;{\rm Kg}\equiv 1.09775\cdot10^{30}\;{\rm n.u.}
\]
\[
1\; {\rm n.u.} [L]=2R_0=\frac{\hbar}{mc}=3.86153\cdot10^{-13}{\rm m},\quad\rightarrow 1\; {\rm m}\equiv 2.58965\cdot10^{12}\;{\rm n.u.}
\]
\[
1\; {\rm n.u.} [T]=\tau_0=\frac{2R_0}{c}=6.44034\cdot10^{-22}\;{\rm s},\quad \rightarrow 1\; {\rm s}\equiv 7.76357\cdot10^{20}\;{\rm n.u.}
\]
\[
e=1 \;{\rm n.u.}\; [Q]=1.6021892\cdot10^{-19} {\rm C},\quad\rightarrow 1 \;{\rm C}\equiv 6.24146\cdot10^{18}\;{\rm n.u.}
\]
The electric field in the International System of units is expressed in V/m. According to the equivalence among units
\begin{equation}
1 \;{\rm V/m}=1\; {\rm m\, Kg\, s^{-2}\,C^{-1}},\quad 1\; {\rm V/m}=7.55676\cdot 10^{-19}\;{\rm n.u.} 
\label{voltM}
\end{equation}
The magnetic field is expressed in teslas. Since
\begin{equation}
1 \;{\rm T}=1\; {\rm Kg\, s^{-1}\,C^{-1}},\quad 1\; {\rm T}=2.26546\cdot 10^{-10}\;{\rm n.u.} 
\label{Teslanu}
\end{equation}

\section{The plane wave}
\label{wave}

\begin{figure}[!hbtp]\centering%
\includegraphics[width=7cm]{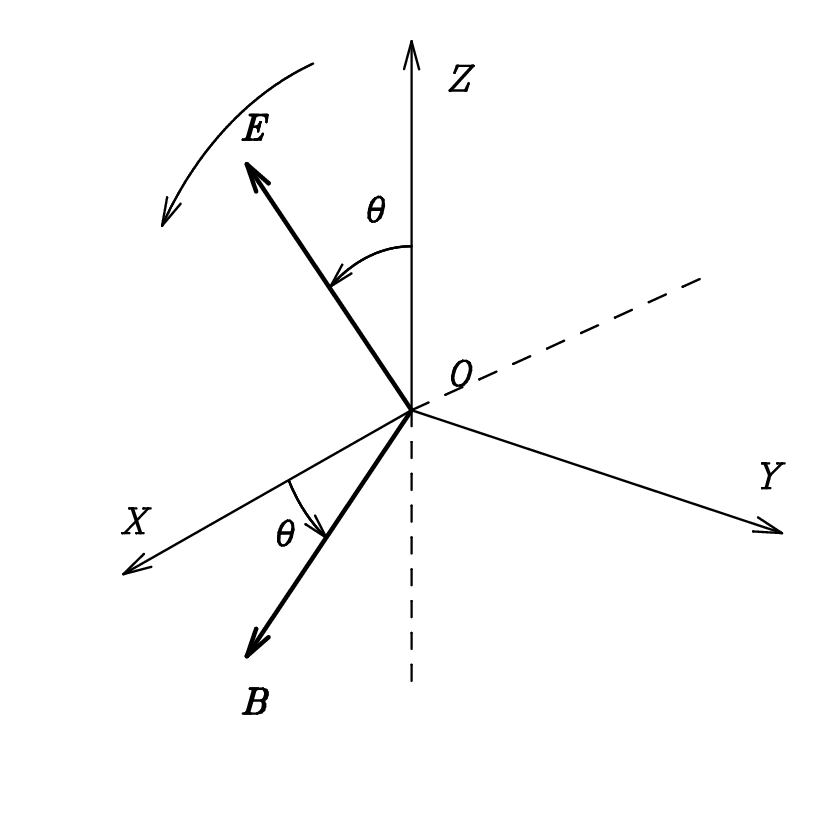}
\caption{Circularly polarized electromagnetic plane wave traveling along the positive $OY$ axis, with the fields on the $XOZ$ plane.
The phase $\theta=\nu (t-y)+\sigma$, in natural units, where $\sigma$ is an arbitrary phase, and the dimensionless $\nu>0$ corresponds to a right handed wave while for  with $\nu<0$ to a left handed circularly polarized wave. } 
\label{fig1:planeWave}
\end{figure}

Let us consider an electromagnetic plane wave, circularly polarized, moving along the $OY$ axis (see figure {\bf\ref{fig1:planeWave}}). The electromagnetic fields of the wave are the rotating orthogonal vectors ${\bi E}$ and ${\bi B}$ of the figure.
The field components of the plane wave are:
\begin{eqnarray}
E_z(t,y)&=&E\cos(\omega (t-y/c)+\sigma),\quad  E_x=k E\sin(\omega (t-y/c)+\sigma),\label{fieldE}\\
B_z(t,y)&=&-k B\sin(\omega (t-y/c)+\sigma),\quad  B_x=B\cos(\omega (t-y/c)+\sigma),
\label{fieldB}
\end{eqnarray}
where $\omega$ is the angular frequency of the wave and $\sigma$ the initial phase. If $\omega>0$ the fields are rotating rightwards when looking along the direction of the motion of the front plane along the $OY$ axis and rotating leftwards if $\omega<0$. The factor $k=1$ and if selected $k=0$ the plane wave is transformed into a linearly polarized plane wave with the electric field oscillating along $OZ$ axis and the magnetic field along $OX$ axis. If the parameter $k$ is in the range $0<k<1$,
the wave corresponds to an elliptically polarized wave with the major axis of the ellipse along the $OX$.
Other orientation of this axis is controlled by changing the relative phases of the two components of the fields $\sigma_x$ and $\sigma_z$. The intensity of both fields in vacuum are related by $E=cB$, in the International System of units. 

In natural units $\widetilde{t}=t/\tau_0$, $\widetilde{y}=y/2R_0$, where $R_0=\hbar/2mc$ is the separation between the CC and the CM at the center of mass frame, and $c=1$, $2R_0=1$ and $\tau_0=2R_0/c=1$, as described in the article \cite{classical1}, and the argument of the sine and cosine functions is
\[
\theta=\omega(\tau_0\widetilde{t}-2R_0\widetilde{y}/c)={\omega}{\tau_0}(\widetilde{t}-\widetilde{y}).
\]
We delete from now on the tildes and the frequency of the wave is the dimensionless parameter $\nu=\omega\tau_0$ and all variables are written in dimensionless natural units.

\section{The dynamical equations}
\label{eqs}
The dynamical equations of the center of mass (CM) ${\bi q}$ and the center of charge (CC) ${\bi r}$ of the Dirac particle under any electromagnetic field in the international system of units are:
\begin{eqnarray}
\frac{d^2{\bi q}}{dt^2}&=&\frac{e}{m\gamma(v)}\left[{\bi E}(t,{\bi r})+{\bi u}\times{\bi B}(t,{\bi r})-\frac{1}{c^2}{\bi v}\left(\left[{\bi E}
+{\bi u}\times{\bi B}\right]\cdot{\bi v}\right)\right],\label{eq:d2qdt2}\\
\frac{d^2{\bi r}}{dt^2}&=&\frac{c^2-{\bi v}\cdot{\bi u}}{({\bi q}-{\bi r})^2}({\bi q}-{\bi r}),\label{eq:d2rdt2}
 \end{eqnarray}
where  ${\bi u}= d{\bi r}/dt$ and  ${\bi v}= d{\bi q}/dt$ are, respectively, the velocities of these two points, with the constraint $|{\bi u}|=c$ and $|{\bi v}|<c$. The external fields are defined at time $t$ at the CC of the particle ${\bi r}$. These dynamical equations do not contain the contribution of the possible radiation-reaction force.

The integration of the equations is done in natural units.
In natural units, the system of differential equations (\ref{eq:d2qdt2}) and (\ref{eq:d2rdt2}) is transformed into the system of first order, once the tildes are deleted,
\begin{equation}
\frac{d{\bi r}}{dt}={\bi u},\quad \frac{d{\bi u}}{dt}=\frac{1-{\bi v}\cdot{\bi u}}{({\bi q}-{\bi r})^2}({\bi q}-{\bi r}),
\label{dineqru}
\end{equation}
\begin{equation}
 \frac{d{\bi q}}{dt}={\bi v},\quad \frac{d{\bi v}}{dt}=\frac{A}{\gamma(v)}[{\bi E}+{\bi u}\times{\bi B}-{\bi v}\left(\left[{\bi E}
+{\bi u}\times{\bi B}\right]\cdot{\bi v}\right)],
\label{dineqqv}
\end{equation}
with the constraints $u=1$ and $v<1$. The fields are in the International System of units and the parameter $A$
\begin{equation}
 A=\frac{e\hbar}{2m^2c^3}=3.778\cdot10^{-19}{\rm m/V},
\label{parameterA}
\end{equation}
where we take for $e$ and $m$ the electric charge and mass of the electron.

Explicitely, the last equation for the interaction with the above circularly polarized electromagnetic plane wave, written separately by components, is
\[
\frac{dv_x}{dt}=\frac{a}{\gamma(v)}\left[(1-u_y -v_x^2 +v_x^2 u_y-v_x v_y u_x )k\sin(\nu(t-r_y)+\sigma)-\right.
\]
\begin{equation}
\left.(v_xv_z-v_xv_zu_y+v_xv_yu_z)\cos(\nu(t-r_y)+\sigma)\right],
\label{dvxdt}
\end{equation}
\[
\frac{dv_y}{dt}=\frac{a}{\gamma(v)}\left[(u_z-v_yv_z+v_yv_zu_y-v_y^2u_z)\cos(\nu(t-r_y)+\sigma)+\right.
\]
\begin{equation}
\left. (u_x-v_yv_x +v_y v_x u_y-v_y^2u_x)k\sin(\nu(t-r_y)+\sigma)\right],
\label{dvydt}
\end{equation}
\[
\frac{dv_z}{dt}=\frac{a}{\gamma(v)}\left[(1-u_y-v_z^2+v_z^2 u_y-v_z v_y u_z)\cos(\nu(t-r_y)+\sigma)-\right.
\]
\begin{equation}
\left.(v_z v_x-v_z v_xu_y+v_z v_yu_x)k\sin(\nu(t-r_y)+\sigma)\right],
\label{dvzdt}
\end{equation}
where the dimensionless parameter $a=AE$, with the electric field $E$ in V/m. If the intensity of the electric field is 1 V/m, the value of the $A$ parameter in natural units is 
\[
A=\frac{e\hbar}{2m^2c^3}=\frac{1}{2}\;{\rm n.u.}
\]
The conversion of V/m of the electric field to natural units is given by the formula (\ref{voltM}) and the dimensionless parameter $a$ for the interaction with an electric field of intensity 1 V/m, is
\[
a=3.778\cdot10^{-19}\;{\rm n.u.}
\]
If the intensity of the electric field is $E$ in V/m units, we have to multiply this factor $a$ by $E$.

The sign of the constant parameter $a$ is the sign of the electric charge of the Dirac particle. For circularly polarized wave $k=1$, and $k=0$ for linear polarization. Please remark that the acceleration of the CM $d{\bi v}/dt$, depends on the values of the fields evaluated at the CC of the particle $r_y$. This acceleration is zero before the front wave has arrived to the location of the CC of the Dirac particle and different from zero since that time onwards, in which we assume the wave packet is still acting on the particle. If the wave packet is of finite extension, the integration is carried out until the end of the packet, making the external force null from that moment onwards when the particle moves freely.

\subsection{Analysis of the dynamical equations }
\label{Analysis}
The dynamical equations (\ref{eq:d2qdt2}) and (\ref{eq:d2rdt2}) are a non-linear system of second-order differential equations depending on several dimensionless constant parameters. We have been unable to find an analytical solution of that system. It is therefore necessary to perform a numerical analysis. This means that we have to integrate the equations for different sets of the parameters, usually called scenarios, and try to obtain conclusions for the different scenarios depending on the physical significance and range of the parameters. A complete mathematical analysis of the above dynamical equations requires an in-depth study of them which is beyond the scope of this work.

\subsection{Physical parameters}
\label{physical}
The physical parameters of the electromagnetic wave and of the dynamical equations are the following dimensionless magnitudes:\\
{\bf Initial phase $\sigma$}: It is the initial phase of the wave $\sigma\in[0,360^\circ]$.\\
{\bf Frequency $\nu$}: It is the frequency of the wave in natural units. It corresponds to the quotient between the physical frequency of the wave $\omega$ and the frequency of the internal motion of the CC around the CM, for the center of mass observer, $\omega_0=2mc^2/\hbar=1.552\cdot 10^{21}$s$^{-1}$. The sign of $\nu>0$ corresponds to a right handed wave and $\nu<0$ to a left handed wave.\\
{\bf Polarization parameter $k$}: Its range is $k\in[0,1]$. For $k=1$ we have a circularly polarized plane wave moving along the positive direction of the $OY$ axis. For $k=0$ it represents a linearly polarized plane wave with the electric field oscillating along the $OZ$ axis. Finally, $0<k<1$, represents an elliptically polarized plane wave with the major axis of the ellipse along the $OX$ axis. Other orientation of this axis is controlled by changing the relative phases of the two components of the fields $\sigma_x$ and $\sigma_z$.\\
{\bf Intensity of the wave $a$}: It is the parameter $a=AE$. If positive corresponds to a positively charged particle and negative in the opposite case.\\
{\bf Boundary variables of the Dirac particle} are the initial spin orientation $\theta$ and $\phi$ and the initial phase $\psi$ of the location of the CC. Finally the absolute value of the CM velocity $v$ and the orientation $\beta$ and $\lambda$ of the velocity vector.

\subsection{Physical examples}
\label{examples}
The frequency of the wave in natural units is the dimensionless parameter $\nu$. For the electron $\omega_0=1.552\cdot 10^{21}$s$^{-1}$. For polarized visible light of average frequency $10^{15}$s$^{-1}$, the constant $\nu\approx 10^{-6}$ and is much smaller for microwaves and radio frequency waves. For electromagnetic ultraviolet radiation, X-rays and high energy $\gamma-$radiation, is greater than this value. To appreciate the influence of this wave at the Dirac particle scale we have to use for $\nu$ a greater value than this because in natural units the time for a turn of the CC of the Dirac particle around the CM is $\pi$, and we have to use integration times longer than $10^6$ units. Some of the integrations with integration time of the order $10^6$, take around one hour in a laptop. For the intensity of the wave, let us see a couple of examples:

\begin{enumerate}

\item{{\bf Visible light}. It is estimated that the average electromagnetic energy coming from the Sun in the upper layers of the atmosphere is around 1350 W/m$^2$. The Poynting vector of a wave is $S=\epsilon_0 cE^2$. The estimated magnitud of the electric and magnetic fields, if assumed a circular polarized wave, is $E=713.15$ V/m and $B=2.37 \mu$T. The average frequency of visible light is $\omega\simeq 10^{15}{\rm s}^{-1}$, so that the parameters for the interaction of an electron with Sun light are:
\[
a=AE\simeq2.7\cdot 10^{-16},\quad \nu=\omega/\omega_0\simeq1.6\cdot 10^{-6}.
\]
These numbers are very low. To depict the motion of the Dirac particle at the scale of natural units in which the separation between the CM and the CC is of order of $10^{-13}$m and at very short times we need to compute the numerical experiment with a higher value of $a$, and longer integration times.}

\item{{\bf High intensity laser}. There exists high intensity lasers of power $10^{19}$W$/{\rm m}^2$. Some of them are pulsed lasers where the duration of the pulse is around $10^{-15}$s$^{-1}$ and $10^{-18}$s$^{-1}$. This corresponds to a polarized wave with an average electric field of intensity $E=6.13\cdot10^{10}$V/m, so that the parameter $a=AE\simeq2.31\cdot10^{-8}$. There is a wide range of frequencies and to appreciate the action on a Dirac particle we have to integrate to very high integration times. With a femptosecond laser this corresponds in natural time to $\omega_0\cdot 10^{-15}\approx 1.55\cdot 10^6$ units. This is the minimum integration time to analyze the interaction of an electron with a femptosecond pulsed laser. After this integration time the particle moves freely.}
\end{enumerate}

\section{Transformation of the CM spin between inertial observers}
\label{Variation}

The expressions of the CC and CM spins for an arbitray inertial observer are (\ref{sCC}) and (\ref{sCM}). The dynamical equations imply that all variables in their definition depend on the CM velocity ${\bi v}$, and the absolute value of both spins will be a function of the center of mass velocity ${\bi v}$, $S(v)$ and $S_{CM}(v)$. From their definition we see that
\[
{\bi v}\cdot{\bi S}({\bi v})={\bi v}\cdot{\bi S}_{CM}({\bi v}),
\]
so that their projections along the CM velocity, sometimes called the hellicity, is the same, although their absolute
value is different. This means that their orientation with respect to the velocity is different. In the free case
they satisfy the dynamical equations:
\[
\frac{d{\bi S}}{dt}={\bi p}\times{\bi u},\quad \frac{d{\bi S}_{CM}}{dt}=0,
\]
where ${\bi p}=\gamma(v){\bi v}$, in natural units. The CM spin is conserved but the CC spin satisfies the same dynamical equation than Dirac's spin operator in the quantum case. The CC spin precess around the direction of the conserved linear momentum, as described in the reference \cite{classical1}.
 
The Pauli-Lubanski four-vector is expressed in terms of the energy $H$, the linear momentum ${\bi p}$ and the CM spin ${\bi S}_{CM}$, $w^\mu\equiv({\bi p}\cdot{\bi S}_{CM},\;H{\bi S}_{CM}/c)$ . As analyzed in Section {\bf 8} and {\bf 8.3} of \cite{{classical1}}, if we call $w^\mu(0)\equiv(0,mc{\bi S}_{CM}(0))$ to the Pauli-Lubanski four-vector for the particle at rest and $w^\mu({\bi v})\equiv(\gamma(v)m{\bi v}\cdot{\bi S}_{CM}({\bi v}),\gamma(v)mc{\bi S}_{CM}({\bi v}))$ to the Pauli-Lubanski four-vector for the particle in the reference frame where the CM is moving at the velocity ${\bi v}$, respectively, the Lorentz transformation $ w^\mu({\bi v})=\Lambda^\mu_\nu({\bi v}) w^\nu(0)$ implies that
\begin{equation}
{\bi S}_{CM}({\bi v})=\frac{1}{\gamma(v)}{\bi S}_{CM}(0)+\frac{\gamma(v)}{(1+\gamma(v))c^2}({\bi v}\cdot{\bi S}_{CM}(0)){\bi v}
\label{vectorS}
\end{equation}
which represents how the CM spin transforms to a moving inertial observer. 
The inverse transformation is
\begin{equation}
{\bi S}_{CM}(0)={\gamma(v)}{\bi S}_{CM}({\bi v})-\frac{\gamma(v)^2}{(1+\gamma(v))c^2}({\bi v}\cdot{\bi S}_{CM}({\bi v})){\bi v}.
\label{vectorinvers}
\end{equation}
The invariant property 
\[
w^\mu({\bi v})w_\mu({\bi v})=w^\mu(0)w_\mu(0)=\gamma(v)^2\left(({\bi v}\cdot{\bi S}_{CM}({\bi v}))^2-{\bi S}_{CM}({\bi v})^2\right)=-{\bi S}_{CM}(0)^2,
\]
and the relation between the absolute values of the spins in two different frames is:
\begin{equation}
S_{CM}(v,\phi)=S_{CM}(0)\sqrt{\frac{1-v^2}{1-v^2\cos^2\phi}},
\label{spinvariation}
\end{equation}
where $\phi$ is the angle between the CM velocity ${\bi v}$ and the direction of the spin ${\bi S}_{CM}({\bi v})$.
In general $v$ and $\phi$, if the particle is not free, are functions of time in that frame, so that the absolute value of the spin is not a constant of the motion.

\begin{figure}[!hbtp]\centering%
 \includegraphics[width=8cm]{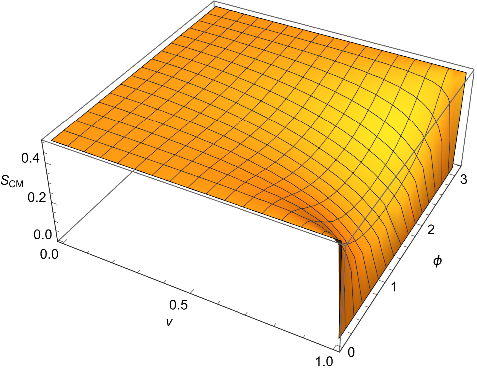}
\caption{Variation of the absolute value of the CM spin with the velocity of the CM and the orientation $\phi$ between the velocity ${\bi v}$ and the center of mass spin ${\bi S}_{CM}({\bi v})$.}  
\label{variation}
\end{figure}

If we take the scalar product of (\ref{vectorS}) with the velocity
\begin{equation}
{\bi v}\cdot{\bi S}_{CM}({\bi v})={\bi v}\cdot{\bi S}_{CM}(0).
\label{vdotS}
\end{equation}
From (\ref{vdotS}) if $\phi_0$ is the angle between ${\bi S}_{CM}(0)$ and ${\bi v}$ it is related to $\phi$ by
\begin{equation}
\cos\phi_0=\cos\phi\,\sqrt{\frac{1-v^2}{1-v^2\cos^2\phi}}.
\label{eq:ficero}
\end{equation}
In natural units $S_{CM}(0)=1/2$, and the variation of $S_{CM}(v,\phi)$ is given in the figure {\bf\ref{variation}}.
From (\ref{spinvariation}) we see that when changing to a moving observer the CM spin changes its absolute value and from (\ref{vectorS}) also its orientation. The first term shows that it shrinks with a $\gamma(v)$ factor and the second term implies that it acquires a component along the relative velocity ${\bi v}$ of order $v^2/2c^2$. From figure {\bf\ref{variation}} we see that the
absolute value of the spin is almost constant at low velocities but decreases at high velocity. If $\phi=0$, $\phi_0=0$, ${\bi v}$, ${\bi S}_{CM}({\bi v})$ and ${\bi S}_{CM}(0)$ are parallel vectors there is no change in the value of the spin, but there is a greater change if ${\bi v}$ and ${\bi S}_{CM}$ are orthogonal, $\phi=\phi_0=\pi/2$, where the behavior is ${\bi S}_{CM}(v)\simeq 1/2\gamma(v)$ and decreases with $\gamma(v)$. Except in the exceptional case that ${\bi v}$ and ${\bi S}_{CM}$ are parallel vectors the value of the CM spin, for an inertial observer who sees the particle at very high velocity, vanishes and the influence of the spin at high energy physics is almost irrelevant.

\subsection{Spin dynamical equations}

To determine how the spin is affected by any external electromagnetic field we have to solve the dynamical equations
(\ref{eq:d2qdt2}) and (\ref{eq:d2rdt2}) and substitute the functions ${\bi r}$, ${\bi q}$, ${\bi u}$ and ${\bi v}$
in the definitions (\ref{sCC}) and (\ref{sCM}), respectively. 

Taking the time derivative of (\ref{sCC}) we get
\[
\frac{d{\bi S}}{dt}=-\frac{d\gamma(v)}{dt}({\bi r}-{\bi q})\times{\bi u}+{\bi p}\times{\bi u},
\]
where
\[
\frac{d\gamma(v)}{dt}=\gamma(v)^3\;{\bi v}\cdot\frac{d{\bi v}}{dt}. 
\]
and leads to:
\[
 \frac{d{\bi S}}{dt}=\gamma(v)^2\left({\bi v}\cdot\frac{d{\bi v}}{dt}\right){\bi S}+{\bi p}\times{\bi u},
\]
From the dynamical equation 
\[
\left({\bi v}\cdot\frac{d{\bi v}}{dt}\right)=\frac{1}{\gamma(v)^3}({\bi v}\cdot{\bi F}),\quad {\bi F}={\bi E}+{\bi u}\times{\bi B},
\]
where ${\bi F}$ is the external Lorentz force. Finally
\begin{equation}
 \frac{d{\bi S}}{dt}=\frac{1}{\gamma(v)}({\bi v}\cdot{\bi F}){\bi S}+{\bi p}\times{\bi u},
\label{spinCCdyn}
\end{equation}
where the term $({\bi v}\cdot{\bi F})$ represents the work, per unit time, of the external Lorentz force along the CM trajectory.

Taking the time derivative of (\ref{sCM}) we get 
\begin{equation}
 \frac{d{\bi S}_{CM}}{dt}= \frac{1}{\gamma(v)}({\bi v}\cdot{\bi F}){\bi S}+({\bi r}-{\bi q})\times{\bi F},
\label{spinCMdyn}
\end{equation}
where the  the first term is the same first term in (\ref{spinCCdyn}) and the last term is the torque of the external Lorentz force defined at the CC ${\bi r}$ with respect to the CM ${\bi q}$.
If the particle is free ${\bi F}=0$, and the spin dynamical equations reduce to $d{\bi S}/dt={\bi p}\times{\bi u}$ and 
$d{\bi S}_{CM}/dt=0$, respectively.

We see that both spins satisfy different dynamical equations. The CC spin ${\bi S}$ precess around the linear momentum ${\bi p}\times{\bi u}$, like Dirac's spin operator (\ref{spinCCdyn}) and it is modified by the work of the external Lorentz force along the CM trajectory and the intensity of this variation is proportional to ${\bi S}$. The CM spin ${\bi S}_{CM}$ is also modified by the interaction with the same work term along the spin ${\bi S}$ and also by a torque of the external Lorentz force (\ref{spinCMdyn}) defined and attached to the CC ${\bi r}$, with respect to the CM ${\bi q}$.

In the case of a cyclotron motion of the CM, the external force produces no work along the CM trajectory and the CC spin ${\bi S}$ satisfies the same dynamical equation than in the free case and the CM spin spin dynamics is the torque with respect to the CM of the external force.

\section{Interaction of a plane wave with a Dirac particle at rest}
\label{Inter}
We are going to consider a Dirac particle at rest in the laboratory with the spin ${\bi S}_{CM}$ oriented with zenithal angle $\theta$ and azimuthal angle $\phi$. The initial phase of the CC is $\psi$. Its CM is located at the point $(d_x,d_y,d_z)$ at time $t=0$ when an electromagnetic packet of consecutive plane waves start from the origin along $OY$ axis at this time. In natural units the velocity of the wave is 1 and the front wave is located at the coordinate $y=t$ at time $t$. The particle is free, the center of mass is at rest and the CC is moving on a plane perpendicular to the spin ${\bi S}_{CM}$, until the front wave reaches the CC of the particle, and the right hand side of equation (\ref{eq:d2qdt2}) becomes different from zero. From now on the wave packet interacts with the Dirac particle that is no longer free and the CM moves. We integrate the dynamical equations with the Mathematica notebook \cite{onda}.
The object of this analysis is to determine the evolution of the CM of the Dirac particle and the spin variation and how the different physical parameters modify the behavior of the particle.

\begin{figure}[!hbtp]\centering%
\includegraphics[width=6cm]{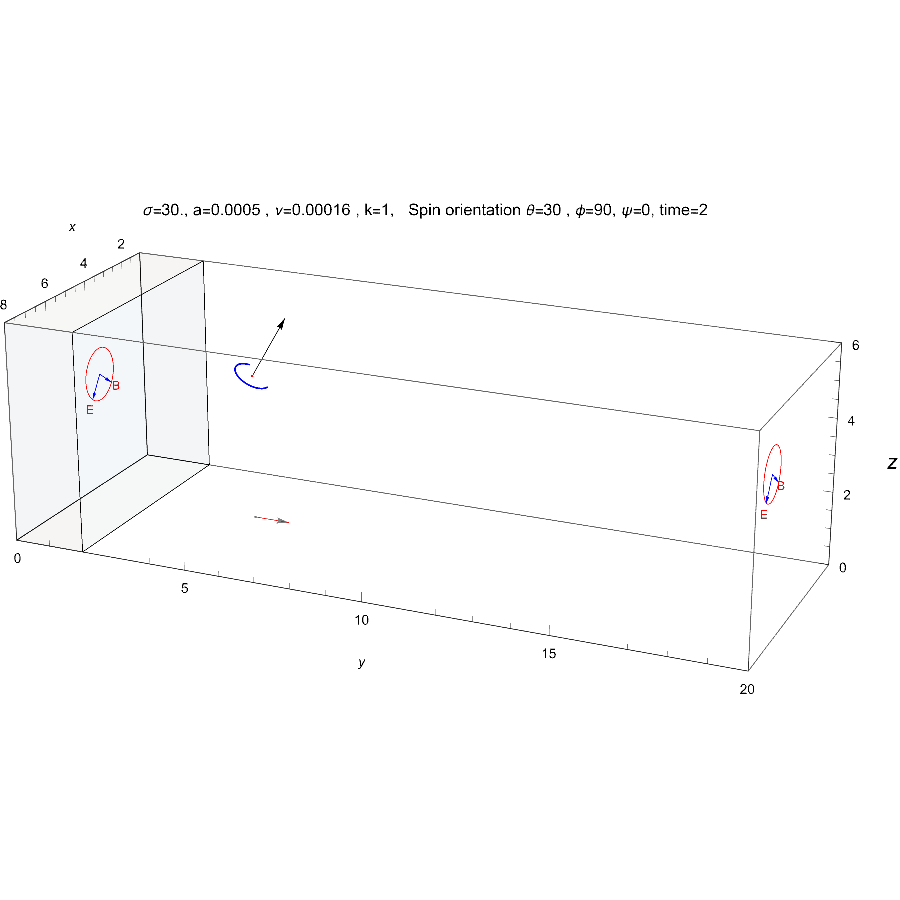}\hspace{0.5cm}
\includegraphics[width=6cm]{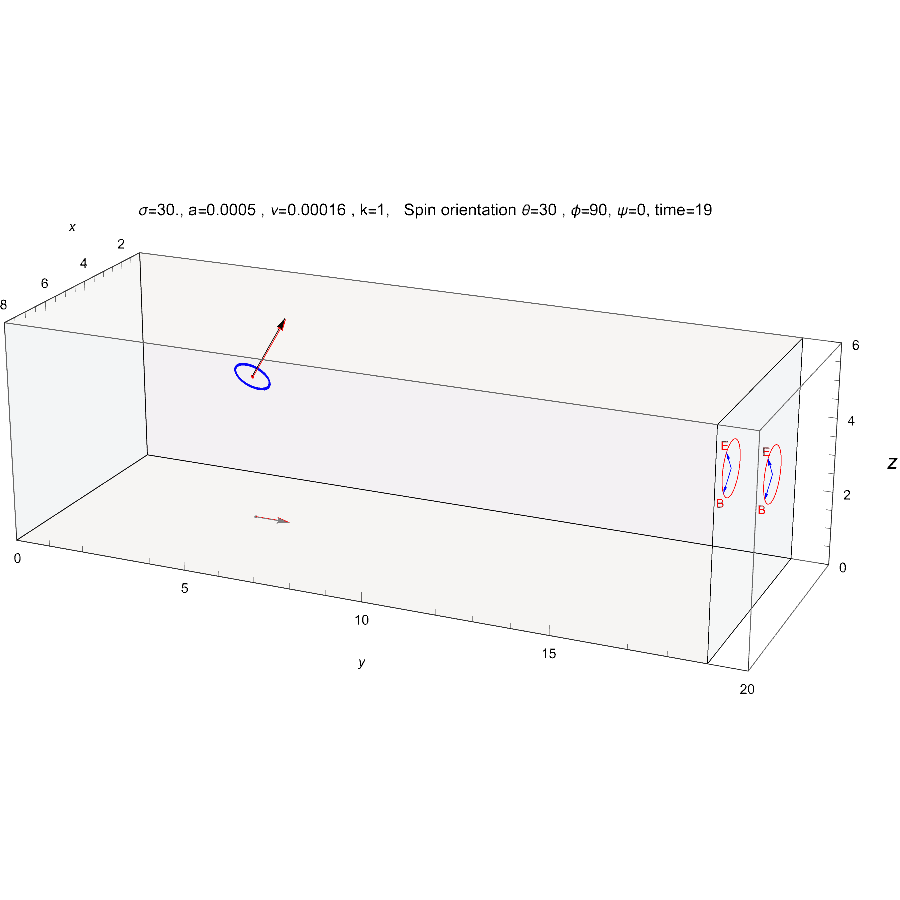}\\
\vspace{-1.4cm}
\includegraphics[width=6cm]{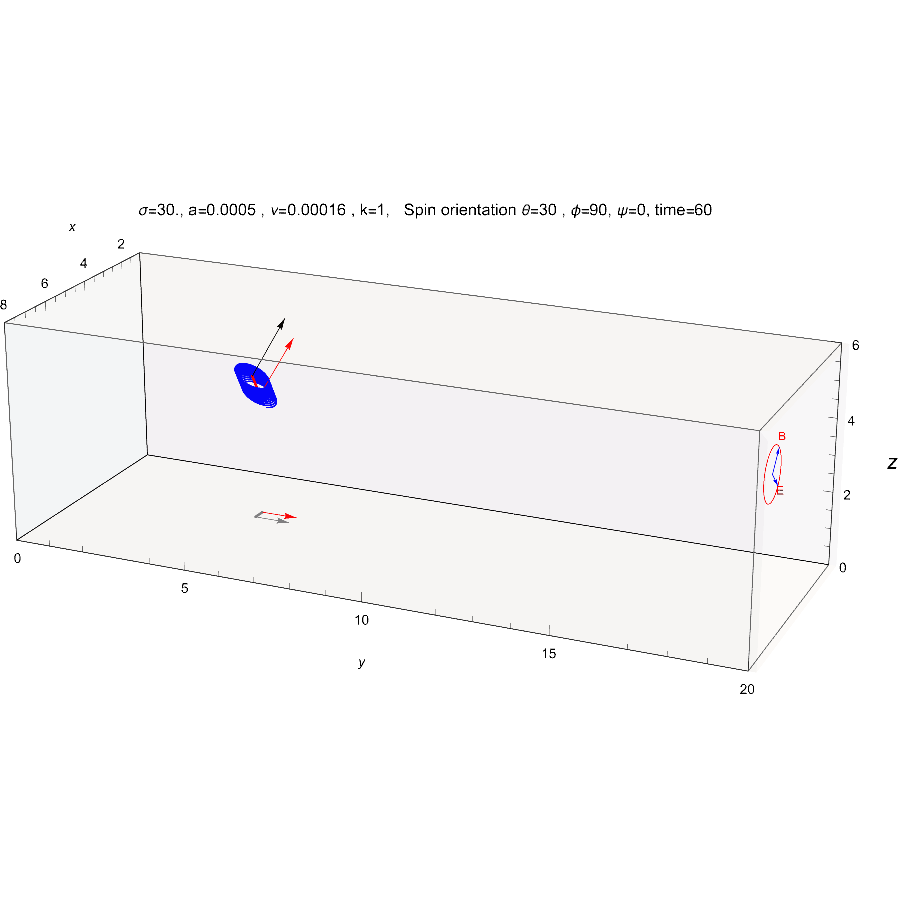}\hspace{0.5cm}
\includegraphics[width=6cm]{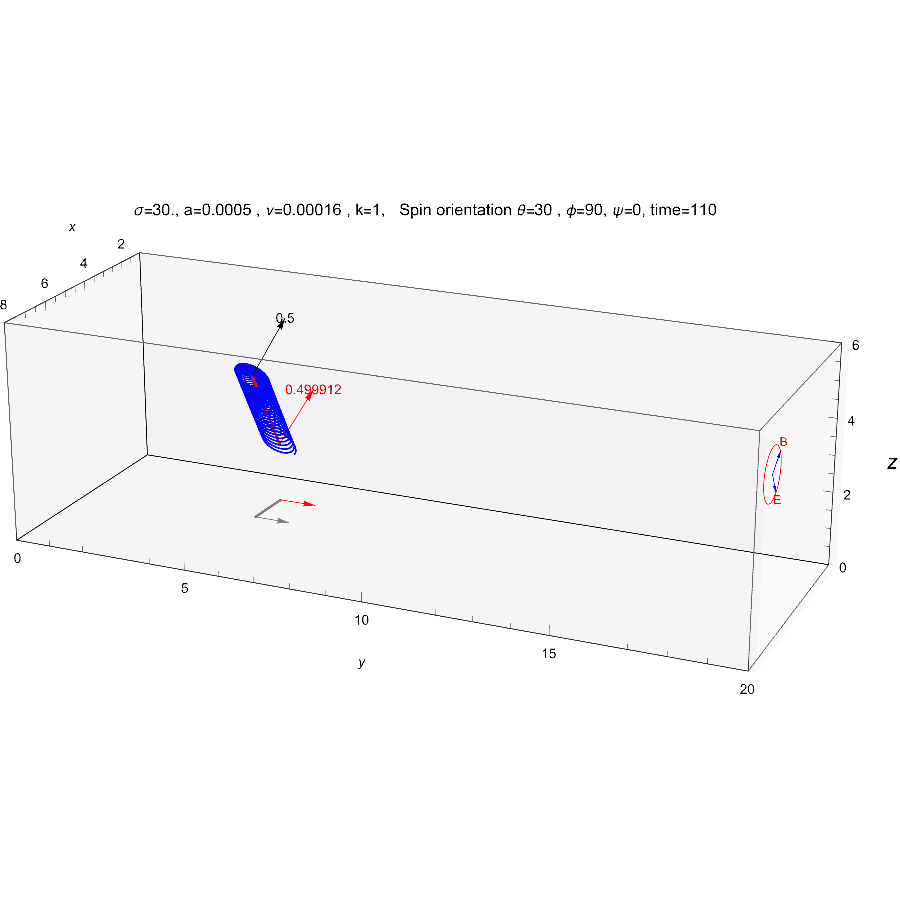}
\caption{The front wave of a circularly polarized plane wave is arriving to a Dirac particle at rest, with the CM located at the point $(4,5,4)$ at $t=0$. The wave starts at time $t=0$ at $y=0$ and the CC of the particle starts moving also at $t=0$. The sequence of pictures corresponds to the instants $t=2, 19, 60$, and $110$ in natural units. The time for a turn of the CC is $\pi$.
During these times the CC of the Dirac particle has rotated respectively almost a three quarters of a turn, around 6 turns, 19 turns and around 35 turns. The CM of the Dirac particle starts moving when the front wave reaches the CC. We also depict the CC and CM motion, the CM spin and the projection of the CM motion and spin on the $XOY$ plane. The initial spin is black and red the spin at the end of integration time. In the front wave it is also depicted the instantaneous orientation of the electric and magnetic fields in that plane as well as in the front end of the picture. For the last two pictures the front wave is out of the screen. The spin orientation is $\theta=30^\circ$ and $\phi=90^\circ$ and $\psi=0$, and we use a big interaction parameter $a=0.0005$. The frequency is $\nu=0.00016$ in natural units and the phase of the field $\sigma=30^\circ$. The electromagnetic wave transfers linear momentum and energy. The linear momentum transfer has a small positive component along the direction of the wave propagation after the integration time. Since the speed acquired by the CM is very small, the orientation and the absolute value of the spin are slightly modified. In the last picture we also depict the absolute values of the initial and final CM spins. The initial spin is $1/2$ and the final spin is $0.499912$, but its projection looks bigger because of a slight change of orientation.}  
\label{fig1:planeWaveVarious}
\end{figure}

The sequence of pictures of figure {\bf\ref{fig1:planeWaveVarious}} shows the front wave at different times before and after reaching the particle. We depict in red the CM motion, in blue the CC motion and in black the CM spin. The red spin vector corresponds to the CM  spin at the end of the integration time. We also depict the projection of these motions and the spin on the $XOY$ plane. The velocity of the wave is 1 in natural units, and the time for the CC revolution is $\pi$ when the particle is at rest and $\pi\gamma(v)$ when moving, also in natural units. The absolute value of the spin in the center of mass frame is $1/2$ in natural units. At this scale the spin vector looks very small and we have depicted the spin vector scaled 4 times in all pictures.
 
\begin{figure}[!hbtp]\centering%
\includegraphics[width=6cm]{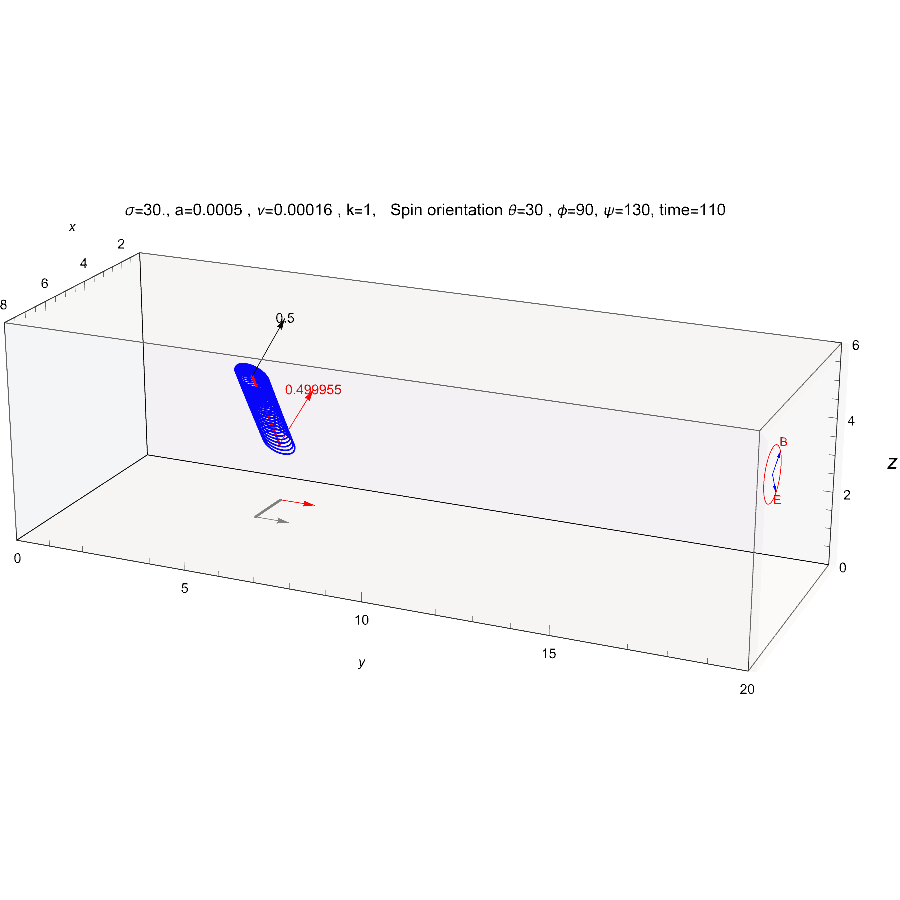}\hspace{0.5cm}
\includegraphics[width=6cm]{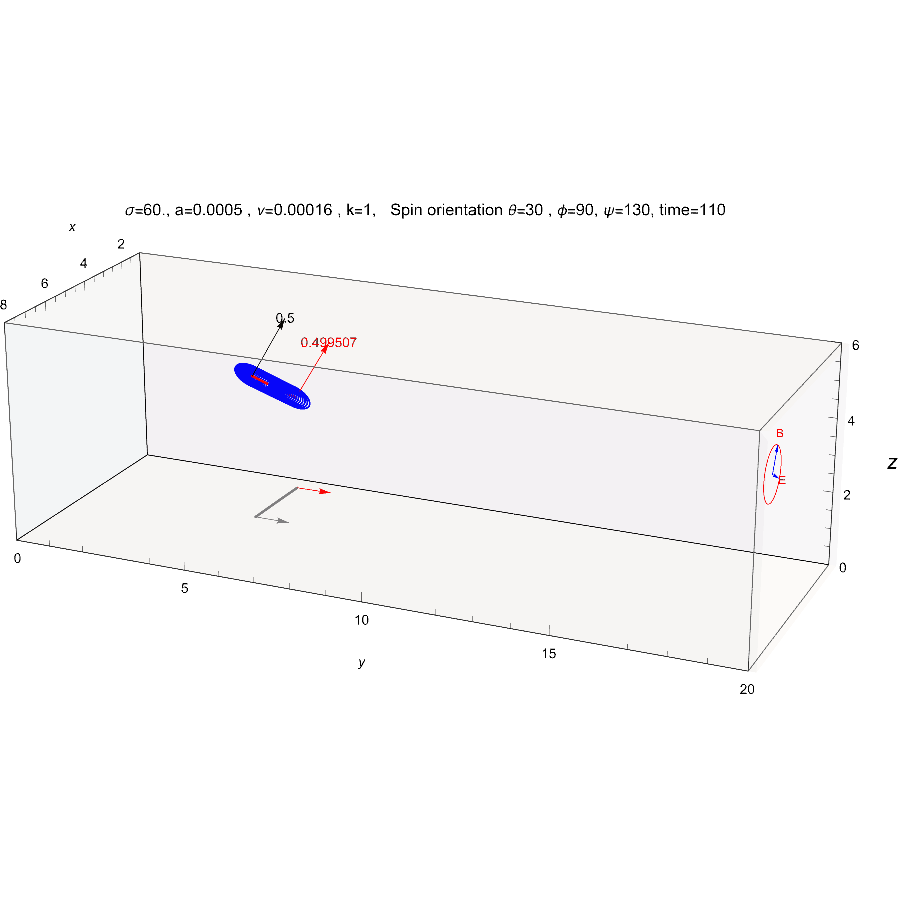}
\caption{The same integration as in figure {\bf\ref{fig1:planeWaveVarious}} but the initial phase of the CC, $\psi=130^\circ$. We find no difference  with the previous picture for the global CM motion although the initial instant and phase of the field when reaches the CC of the Dirac particle is different. The right picture the initial phase is $\sigma=60^\circ$ and the CM trajectory has rotated around $30^\circ$.}  
\label{fig1:planeWaveVariousPsi}
\end{figure}

\begin{figure}[!hbtp]\centering%
 \includegraphics[width=7cm]{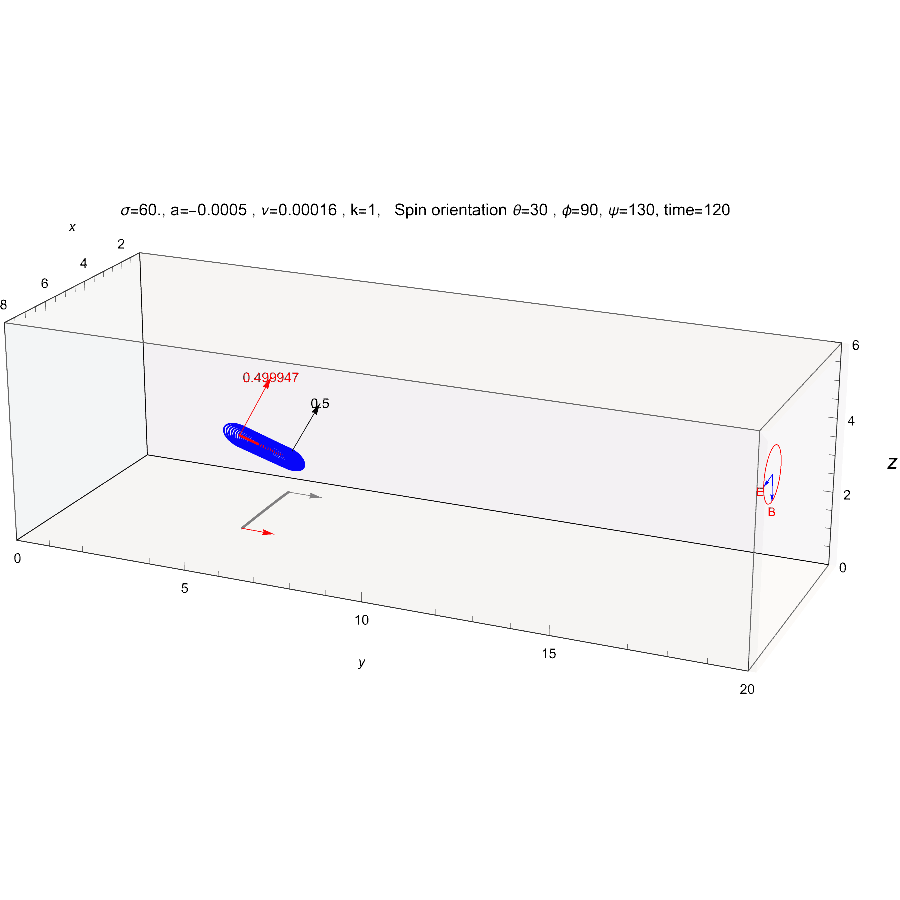}
\caption{The same calculation of the Dirac particle but of opposite electric charge under the same electromagnetic plane wave as in the figure {\bf\ref{fig1:planeWaveVariousPsi}}. The CM of the particle is located inititally at $(2,5,1)$. It moves in the opposite direction than the particle. The electromagnetic wave transfers linear momentum and energy and modifies slightly the absolute value and the CM spin orientation of the antiparticle. The linear momentum transfer has also a positive component along the direction of the wave propagation.}  
\label{fig1:planeWaveanti}
\end{figure}

The above interaction is repeated in the figure {\bf\ref{fig1:planeWaveVariousPsi}} where the only change is the initial phase of the CC, $\psi=130^\circ$. It seems that this change does not affect the global motion of the CM of particle, although the instant and phase when the wave reaches the CC is different. This phase only affects the motion in the early stages as we shall analyze in section {\bf\ref{point}} when comparing this motion with the interaction with a point particle. The right figure is the same integration with the phase of the field $\sigma$ changed from $30^\circ$ to $60^\circ$. This is what basically the trajectory of the CM has rotated.\\
In figure {\bf\ref{fig1:planeWaveanti}} we repeat the calculation with a Dirac particle with the same spin orientation but of opposite electric charge thus showing that the antiparticle, like the particle, is moving with a small component of its linear momentum along the wave motion. The absolute value of the spin decreases and the final spin projection looks small because the spin orientation has changed.
The global motion of the CM depends on the phase of the fields $\sigma$ when arriving to the position of the CC of the particle and the initial phase of the CC $\psi$ is only relevant at the early stages of the interaction but it has no influence in the overall CM motion as we shall see later.

\begin{figure}[!hbtp]\centering%
 \includegraphics[width=8cm]{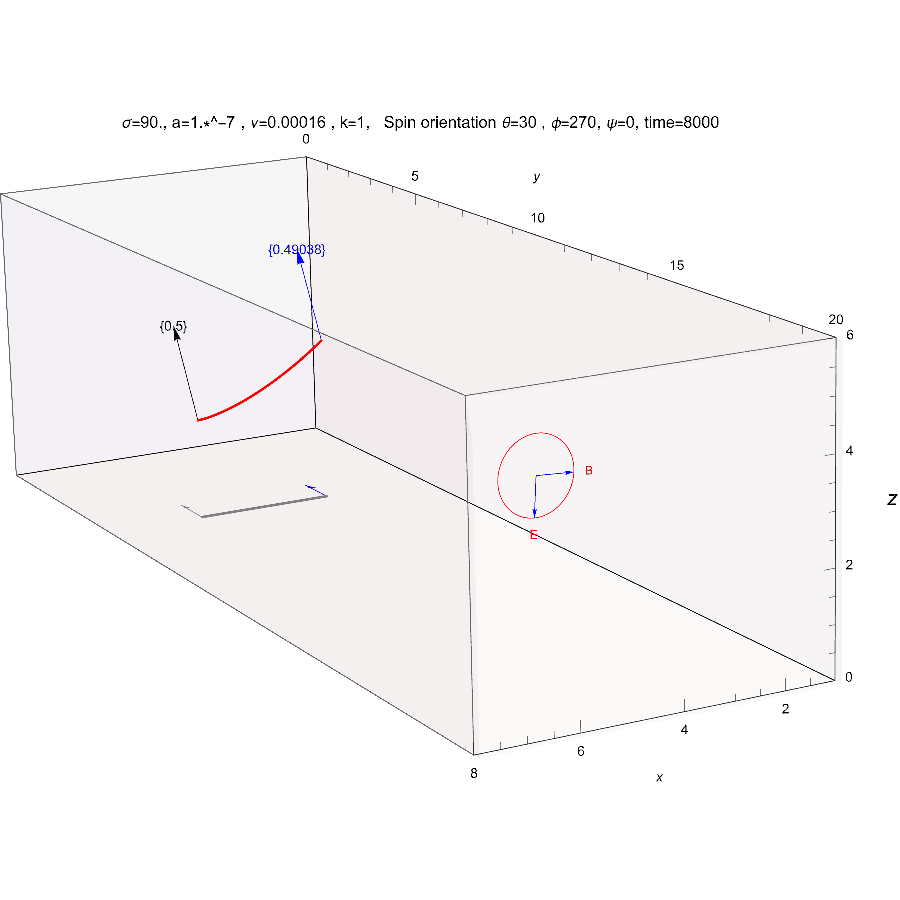}
\caption{Motion of the CM of the Dirac particle at rest (red) under an electromagnetic wave as in the previous figures of wave parameters $a=10^{-7}$, $\sigma=90^\circ$, $\nu=0.00016$, $k=1$ and integration time $t=8000$ natural units. The spin orientation is $\theta=30^\circ$ and $\phi=270^\circ$. The CM has been displaced around $3$ natural units. The motion of the CC has given 2560 turns and hides or blurs the CM trajectory and is not depicted. The last spin is depicted in blue to distinguish from the CM trajectory.
The average final speed acquired by the electron is $v/c\simeq3/8000=0.000375$.  }  
\label{fig:a_7}
\end{figure}

\newpage
To analyze the interaction with a very high energy laser we perform the integration of the Dirac particle, shown in the figure {\bf\ref{fig:a_7}} with the constant $a=1\cdot10^{-7}$ during an integration time of value $t=8000$ natural units. This corresponds to around
2560 turns of the CC around the CM, and $t=5.15\cdot10^{-18}$s of the physical time of the electron. During this time the CM has been displaced around a distance of around $3$ natural units, so that the final CM velocity is $v\approx 0.000375$ and the final spin is of absolute value $S_{CM}=0.49038$. 

\begin{figure}[!hbtp]\centering%
 \includegraphics[width=5cm]{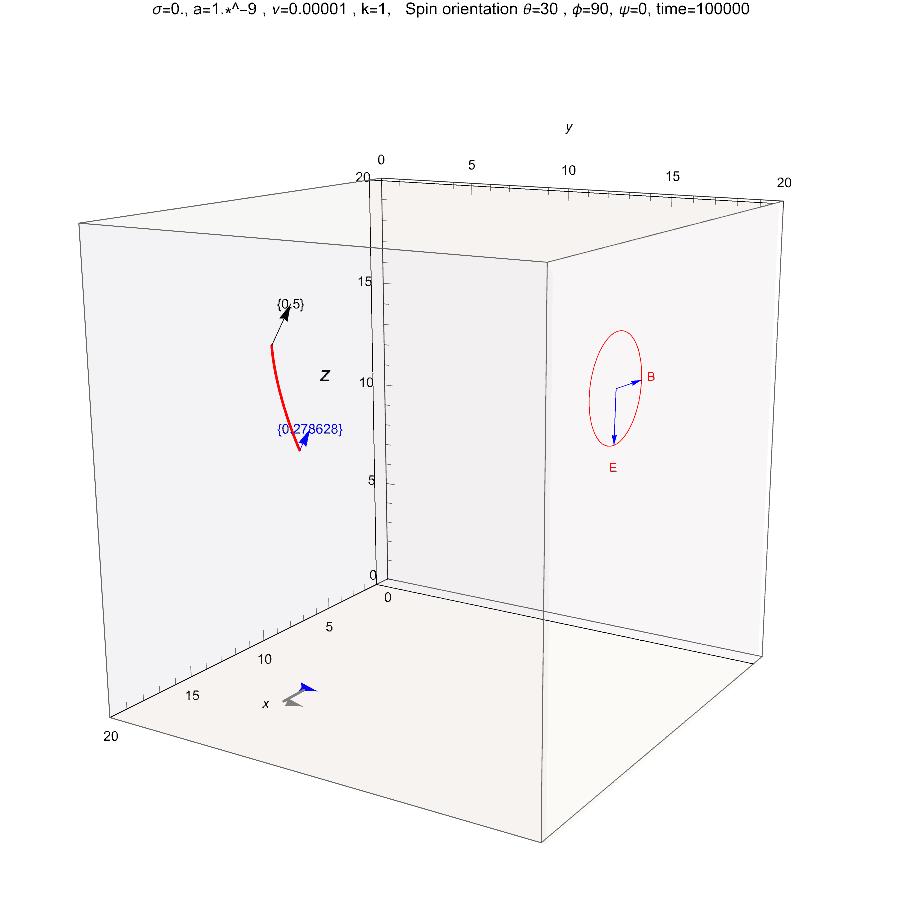} \includegraphics[width=5cm]{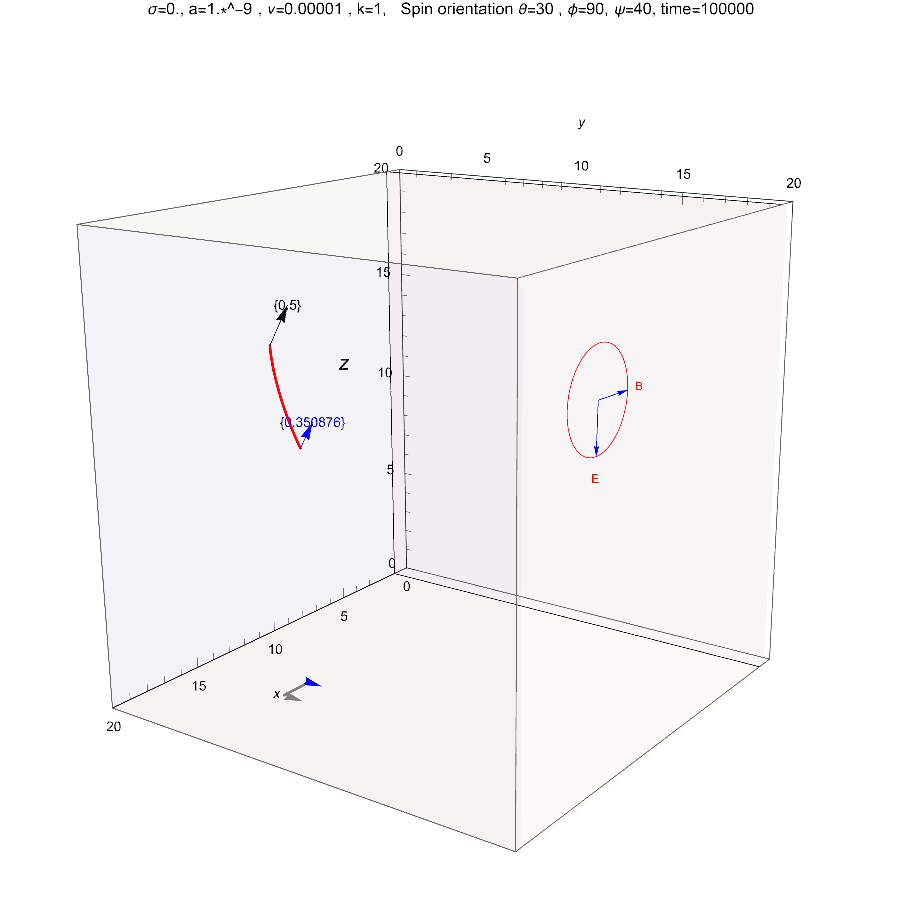}\includegraphics[width=5cm]{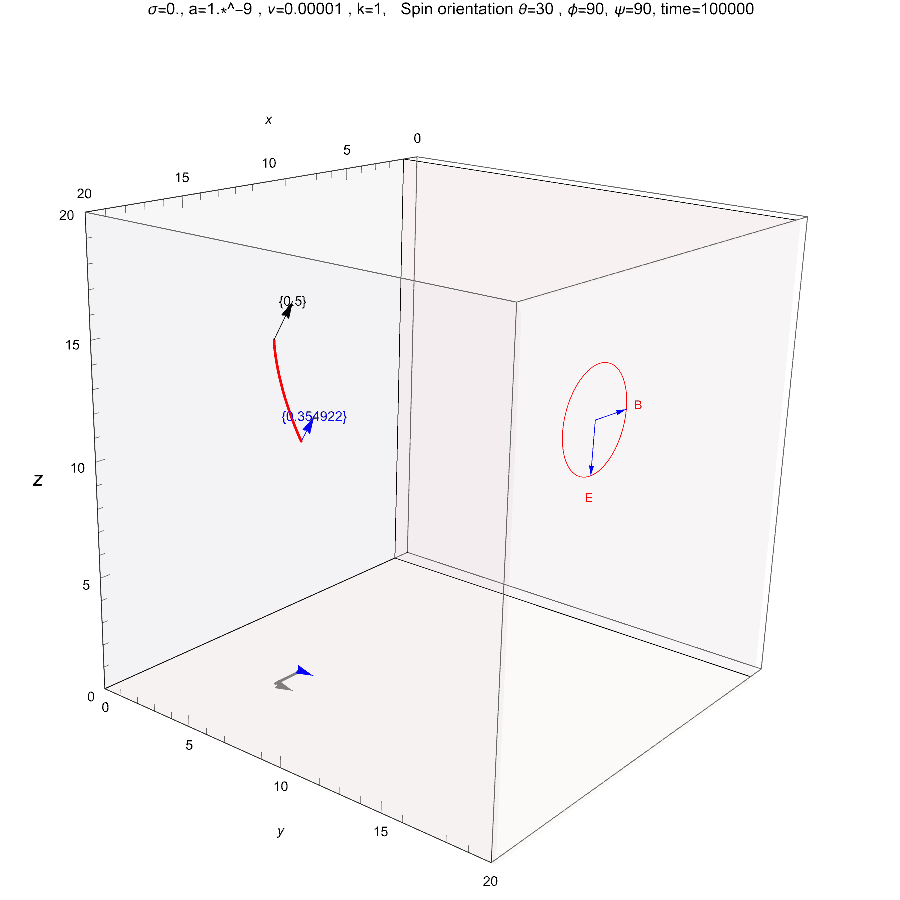}
\caption{Motion of the CM with $a=10^{-9}$, $\nu=0.00001$, $\sigma=0^\circ$, integration time $t=10^5$ and the initial phases of the CC are $\psi=0^\circ$, $40^\circ$ and $90^\circ$, respectively. The initial phase $\psi$ of the CC seems to produce no difference for long integration times. As we shall see this phase only produces a slight modification of the CM trajectory when the CM starts moving and has no influence in the subsequent integration.}  
\label{fig:a_9psi90}
\end{figure}

\begin{figure}[!hbtp]\centering%
\includegraphics[width=6cm]{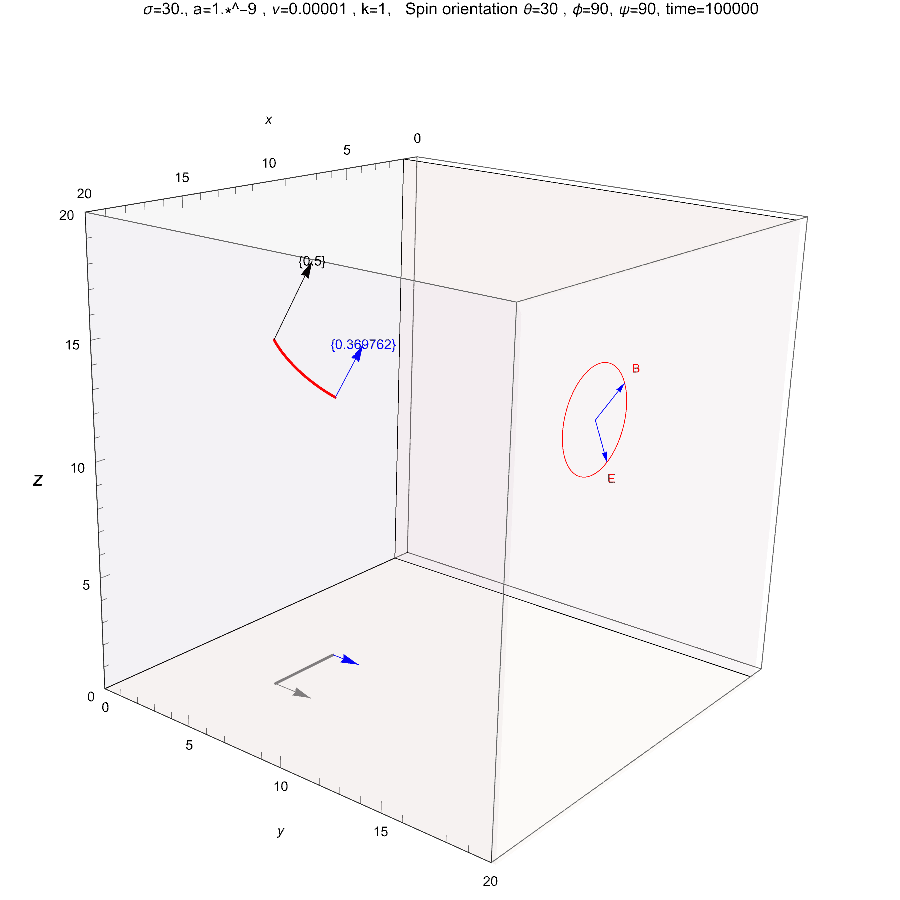}\includegraphics[width=6cm]{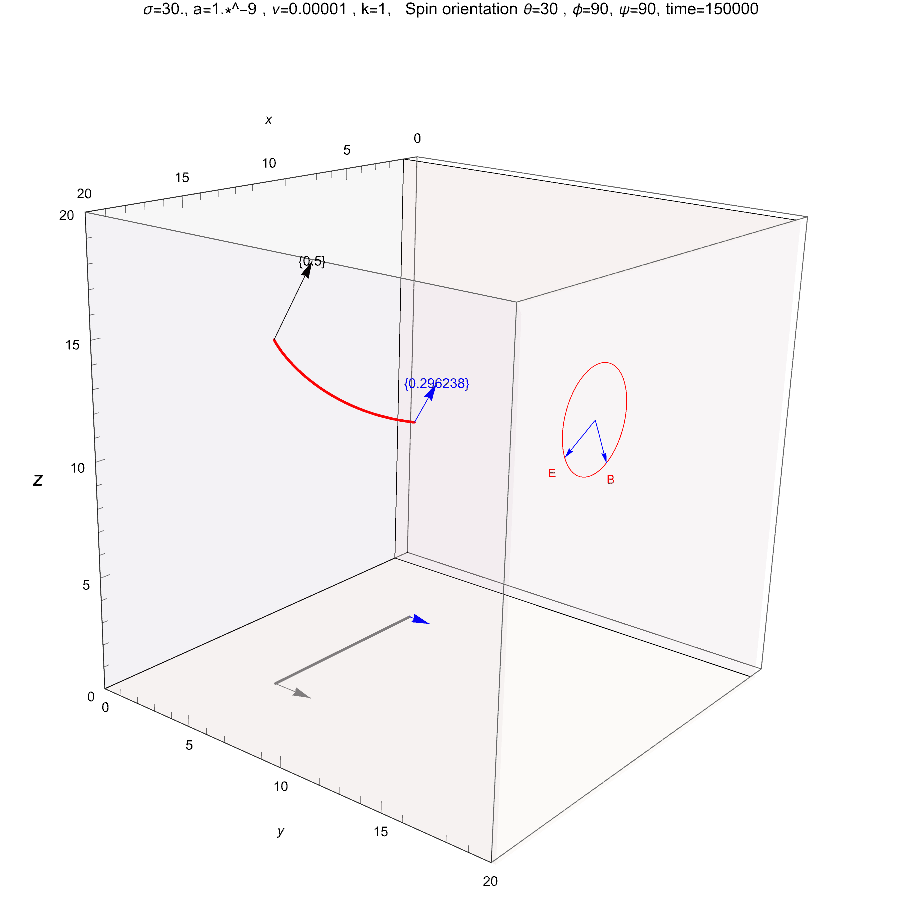}
\caption{Motion of the CM with $a=10^{-9}$, $\nu=0.00001$ and the initial phase of the CC is $\psi=90^\circ$, and $\sigma=30^\circ$ for two different integration times  $t_1=10^5$ and $t_2=1.5\cdot10^5$. The spin has been modified slightly its orientation, but the absolute value has decreased because of the increase of the CM velocity which is almost orthogonal to the spin direction, as discussed in the Section {\bf\ref{Variation}}. The scale of the spin vector in this figure has been enlarged twice with respect to the previous pictures.}  
\label{fig:a_9psi90s30}
\end{figure}

To show how the internal phase $\psi$ of the CC modifies the motion of the CM, let us see the following two figures {\bf\ref{fig:a_9psi90}} and {\bf\ref{fig:a_9psi90s30}}, with the same electromagnetic plane wave but for longer integration times. 
\newpage
\subsection{Interaction with a linearly polarized plane wave}
\label{linearpol}
\begin{figure}[!hbtp]\centering%
\includegraphics[width=5cm]{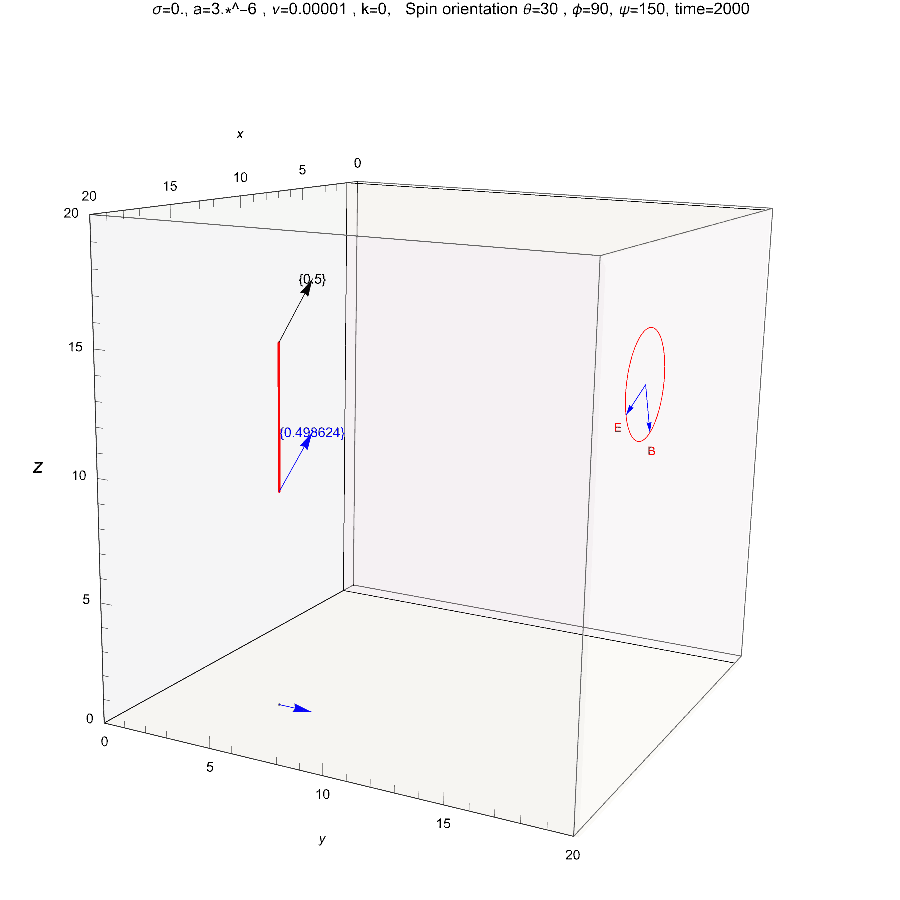}
\includegraphics[width=5cm]{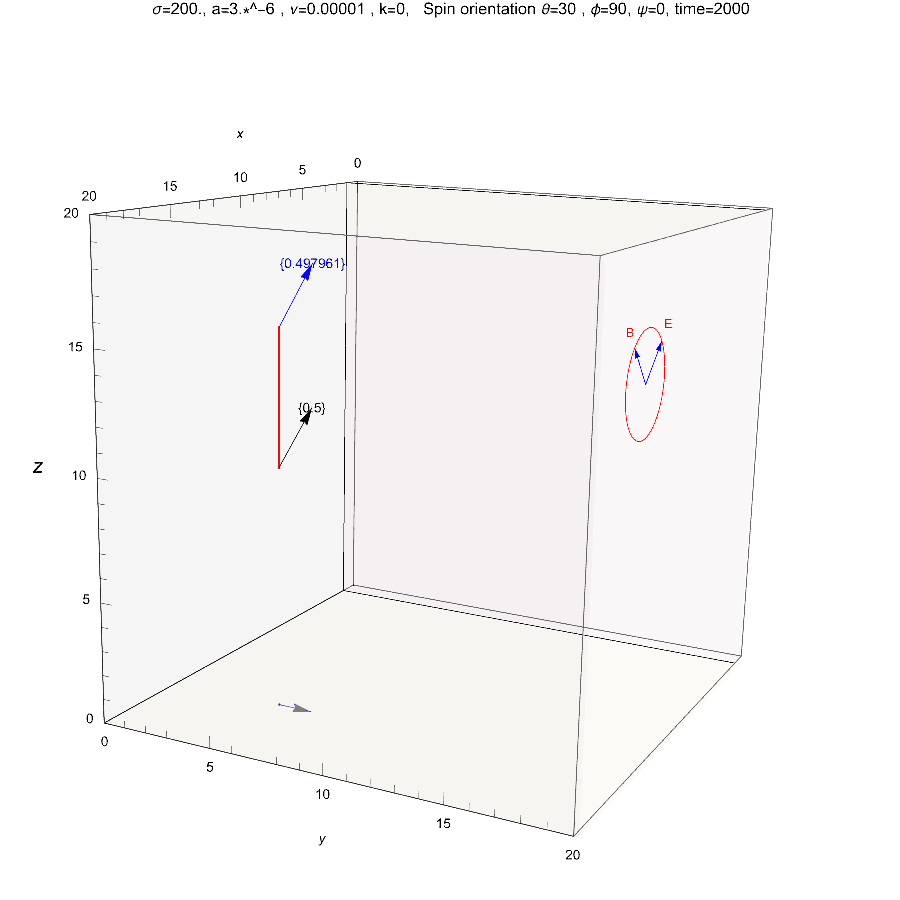}
\includegraphics[width=5cm]{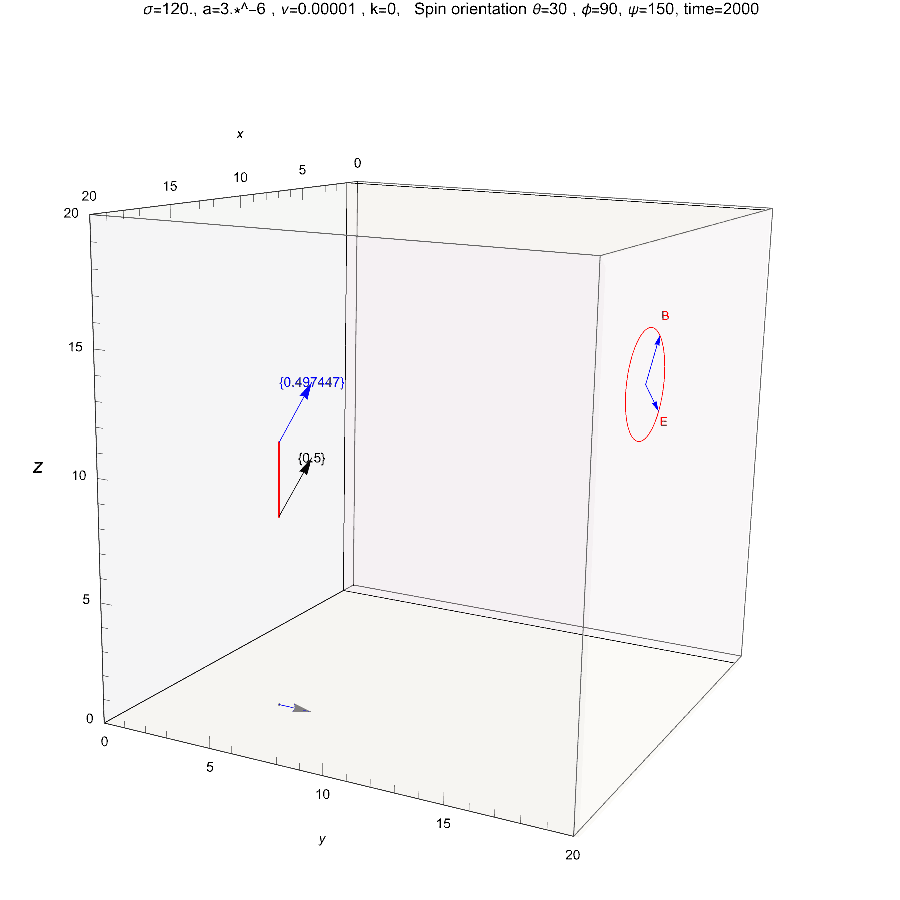}
\includegraphics[width=5cm]{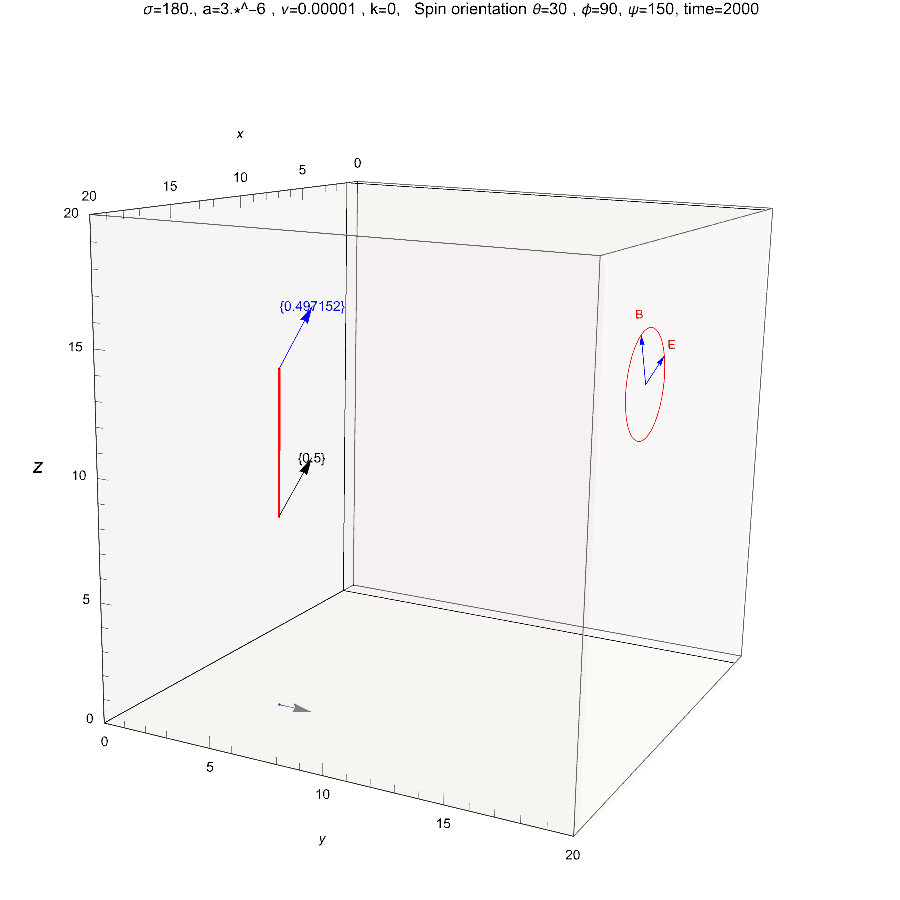}
\includegraphics[width=5cm]{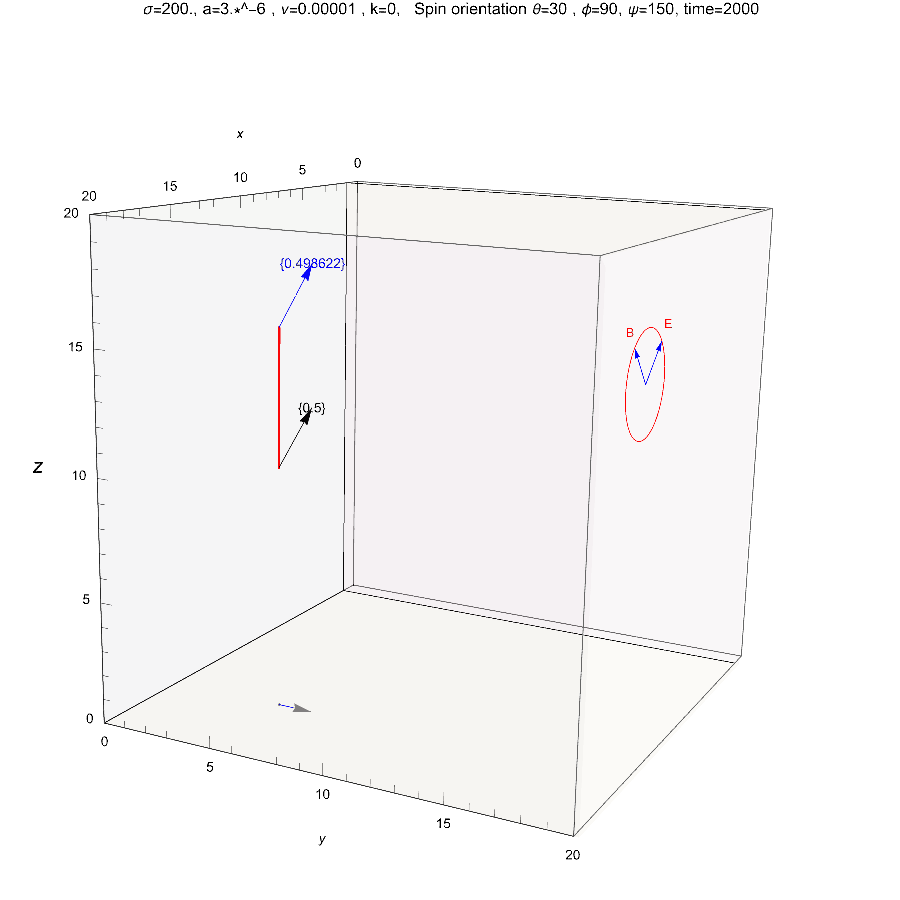}
\includegraphics[width=5cm]{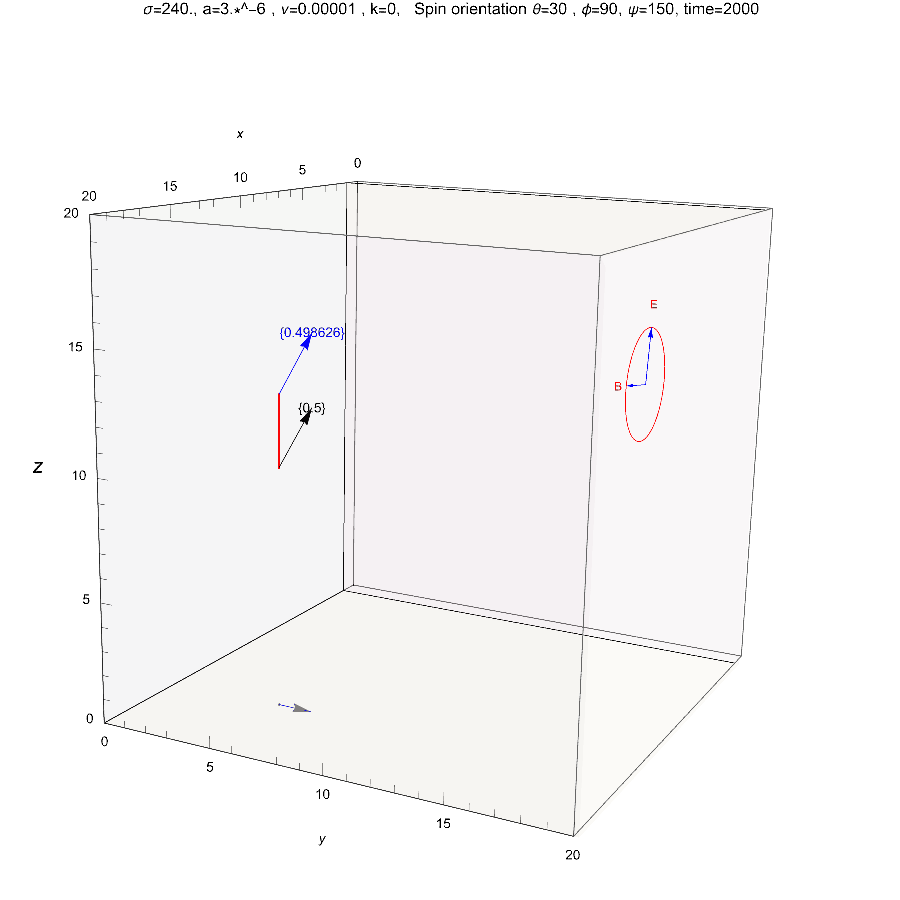}
\includegraphics[width=5cm]{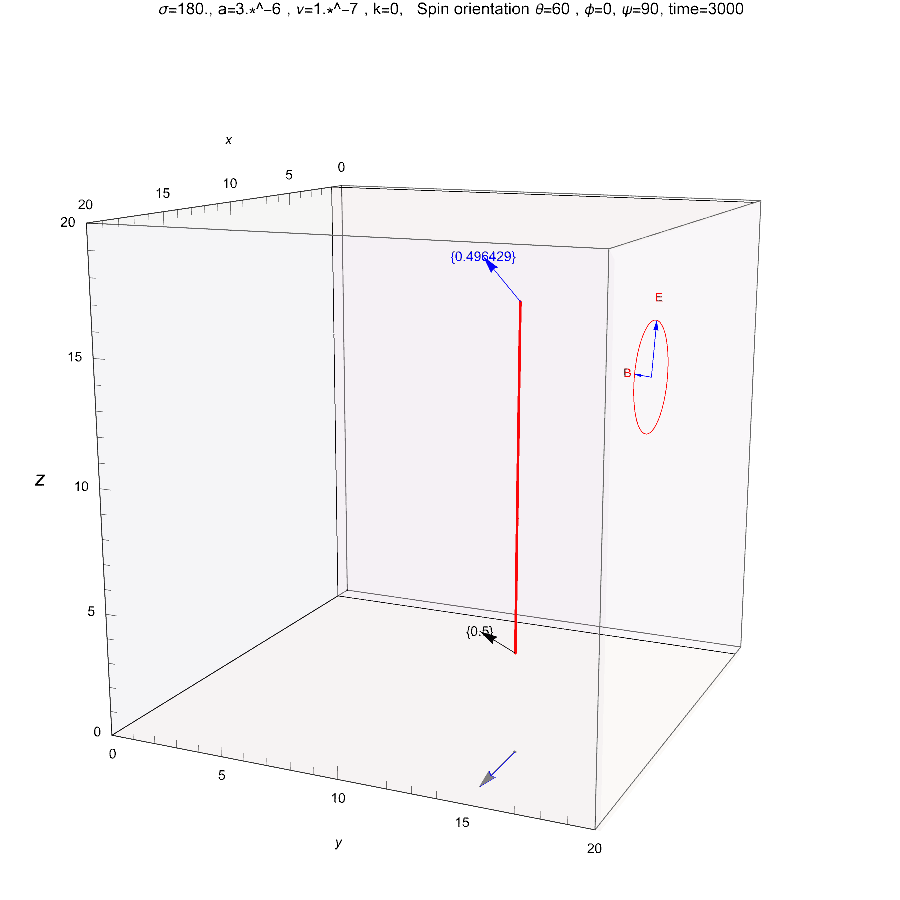}
\includegraphics[width=5cm]{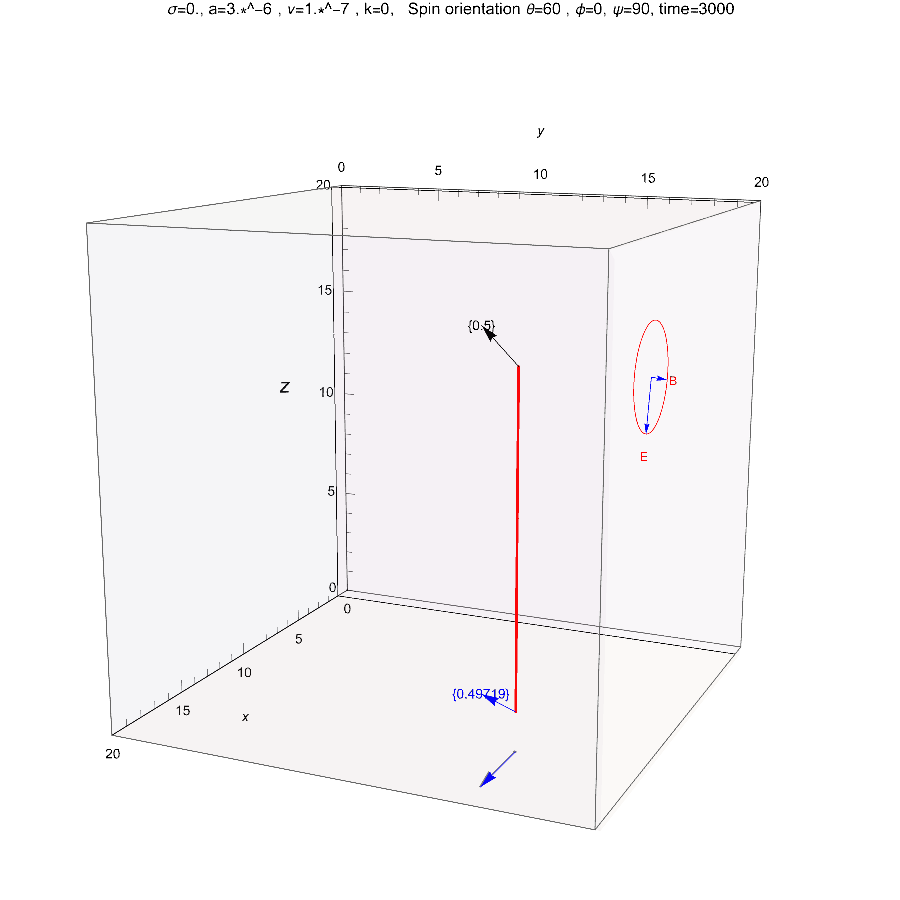}
\includegraphics[width=5cm]{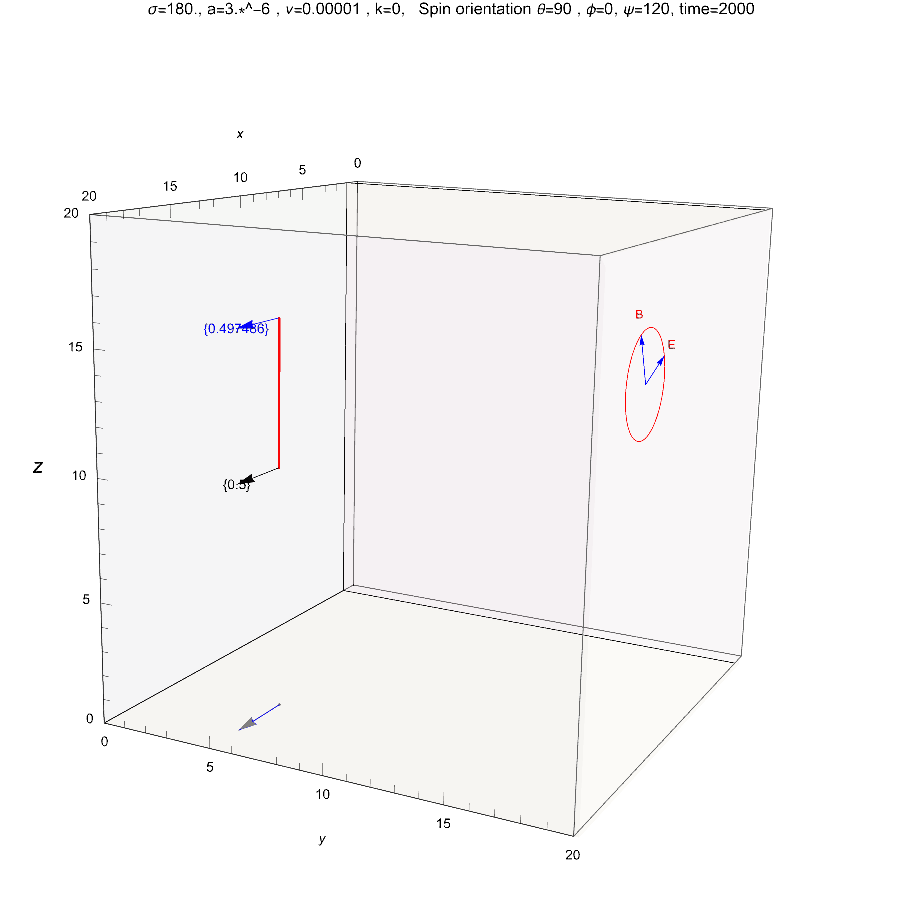}
\includegraphics[width=5cm]{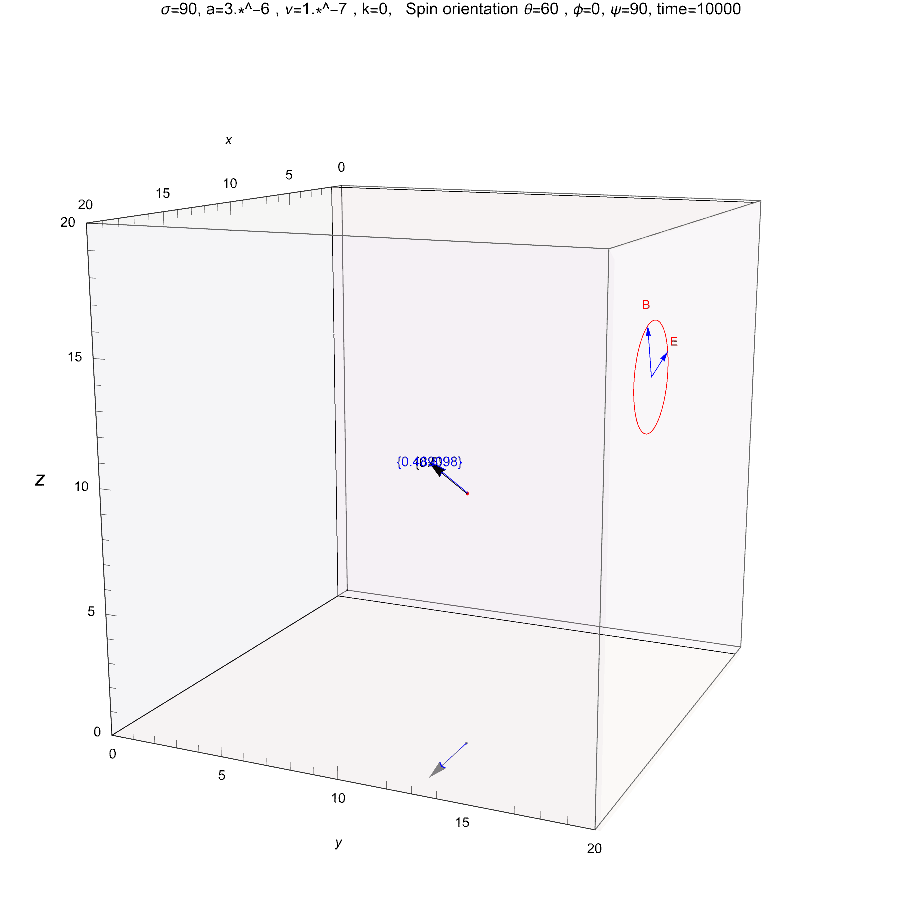}
\includegraphics[width=5cm]{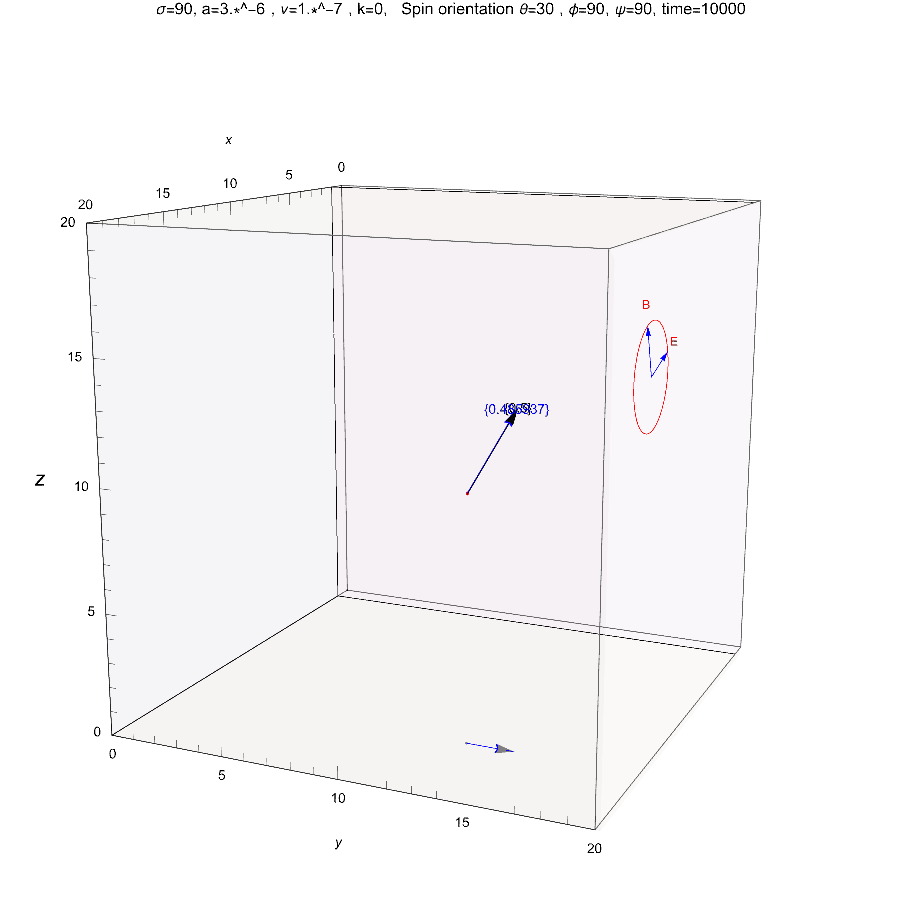}
\includegraphics[width=5cm]{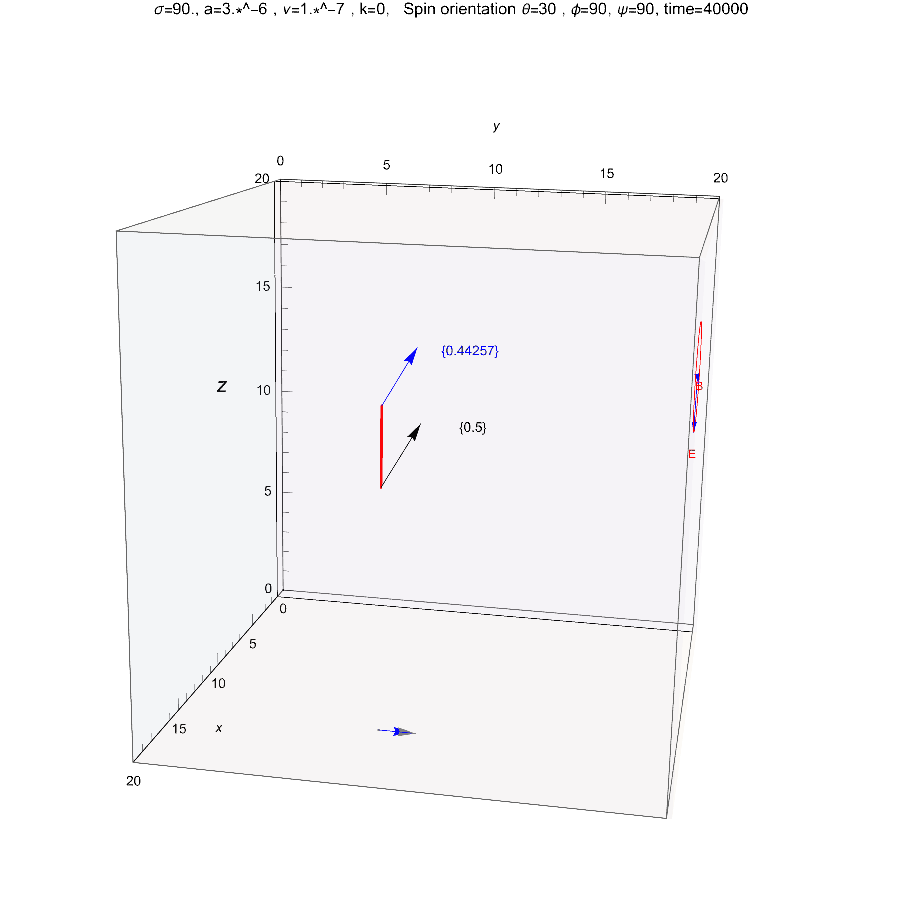}
\caption{Interaction of the Dirac particle with a linear polarized plane wave for different values of the physical parameters. The electric field of the wave oscillates along OZ axis, and it is in this direction where the CM of the Dirac particle is moving. The pictures at bottom, are computed with integration time of $10^4$ units, and for $\sigma=90$ the plane wave do not modify the location of the particle. Nevertheless the last integration is the same as the previous one but with a longer integration time of $t=4\cdot10^4$ with a nonvanishing displacement. A longer integration time produces a linear momentum transfer in the direction of the wave propagation.}  
\label{fig:linearpolwave}
\end{figure}

If we select the parameter $k=0$ we analyze the interaction of the Dirac particle at rest with a linearly polarized electromagnetic plane wave. We depict in the figure {\bf\ref{fig:linearpolwave}} several interactions.
The electric field is oscillating along OZ axis and this is the direction of the displacement of the CM in the upward or downward direction according to the value of the parameter $\sigma$. Upward is for $\sigma\in[90^\circ,270^\circ]$ while downward is produced for $\sigma\in[-90^\circ,90^\circ]$. Most of the integrations are for a time of 2000 units, and different spin orientations are considered. The length of the path depends on the value of $\sigma$ and the greatest displacements are performed for $\sigma=0$ or $180^\circ$. and this is independent of the time of arrival of the front wave to interact with the particle, which is determined by changing the location of the initial position of the CM of the particle.

The last three pictures at bottom show that no displacement of the CM happens if the field parameter $\sigma=90$, or $270^\circ$. The examples are for an integration time of $10^4$ units, for different spin orientations and different CC phase and position of the CM.

\newpage
\section{Interaction of a plane wave with a point particle at rest}
\label{point}

We are going to compare the motion of the Dirac particle with the motion of the spinless point particle (PP) under the same electromagnetic plane wave and in the natural system of units. In this case ${\bi q}_p={\bi r}_p$, ${\bi v}_p={\bi u}_p$, and the dynamical equations for the point ${\bi q}_p$, which represents both the CC and the CM of the point particle are
\[
\frac{d{\bi p}_p}{dt}=e({\bi E}+{\bi v}_p\times{\bi B}).
\]
For the center of mass acceleration of the point particle the equation is
\begin{equation}
\frac{d^2{\bi q}_p}{dt^2}=\frac{e}{m\gamma(v_p)}[{\bi E}+{\bi v}_p\times{\bi B}-\frac{{\bi v}_p}{c^2}({\bi v}_p\cdot{\bi E})].
\label{eqCMPP}
\end{equation}
If we use the same system of natural units the dynamical equations to be solved are the first order system:
\[
\frac{d{\bi q}_p}{dt}={\bi v}_p,\quad \frac{d{\bi v}_p}{dt}=\frac{e}{m\gamma(v_p)}[{\bi E}+{\bi v}_p\times{\bi B}-\frac{{\bi v}_p}{c^2}({\bi v}_p\cdot{\bi E})].
\]
and this last equation is explicitely given by:
\begin{eqnarray}
\frac{dv_{px}}{dt}&=&\frac{a}{\gamma(v_p)}\left[(1-v_{py} - v_{px}^2)k\sin(\nu(t-q_{py})+\sigma)-\right.\nonumber\\&&\left.v_{px}v_{pz}\cos(\nu(t-q_{py})+\sigma)\right],
\label{dvxdtpoint}
\end{eqnarray}
\begin{eqnarray}
\frac{dv_{py}}{dt}&=&\frac{a}{\gamma(v_p)}\left[(v_{pz}-v_{py}v_{pz})\cos(\nu(t-q_{py})+\sigma)+\right.\nonumber\\&&\left.(v_{px}-v_{px}v_{py})k\sin(\nu(t-q_{py})+\sigma)\right],
\label{dvydtpoint}
\end{eqnarray}
\begin{eqnarray}
\frac{dv_{pz}}{dt}&=&\frac{a}{\gamma(v_p)}\left[(1-v_{py}-v_{pz}^2)\cos(\nu(t-q_{py})+\sigma)-\right.\nonumber\\&&\left.v_{px} v_{pz} k\sin(\nu(t-q_{py})+\sigma)\right],
\label{dvzdtpoint}
\end{eqnarray}
where the parameter $a=AE$ is the same as before. The boundary conditions for the point particle will be the same as the boundary conditions for the CM of the Dirac particle.

At first glance we see that if the Dirac and point particle are at rest, inspection of equations (\ref{eq:d2qdt2}) and (\ref{eqCMPP}) show that at the initial interaction time the magnetic force terms are different. Since ${\bi v}_p=0$, there is no magnetic force on the point particle while the ${\bi u}\times{\bi B}$ term of the Dirac particle with $u=c$, is of the same order of magnitude like the ${\bi E}$ term. In addition to this, depending on the spin orientation  and the internal phase $\psi$, the front wave can reach the CC of the Dirac particle before or after the arrival of the wave to the CM and to the point particle, located initially at the CM of the Dirac particle, so that the CM's of the two particles will start moving at different times.
As we shall see this interpretation is correct but the physical effect is that this difference will be detected only at the early stages of evolution and for high intensity waves, while for normal or low energy electromagnetic waves and very long integration time this difference will be negligible. We make the integrations by considering that the evolution parameter $a$ is positive, like the charge of the particles.

\begin{figure}[!hbtp]\centering%
 \includegraphics[width=5cm]{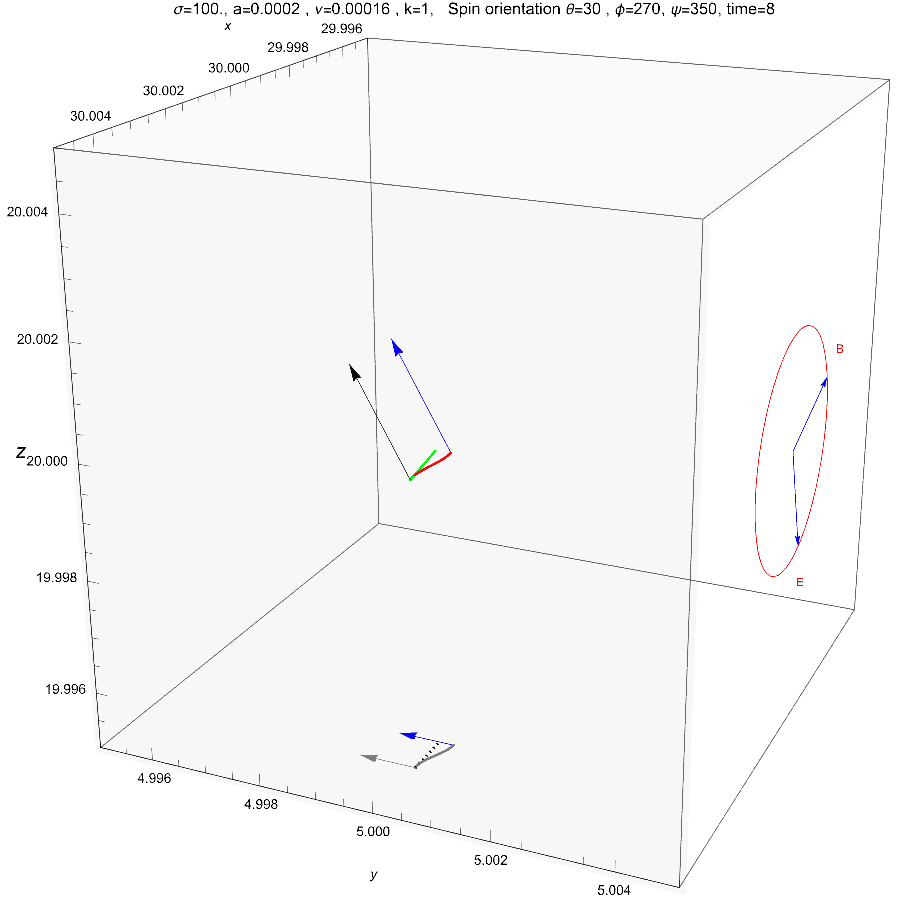} \includegraphics[width=5cm]{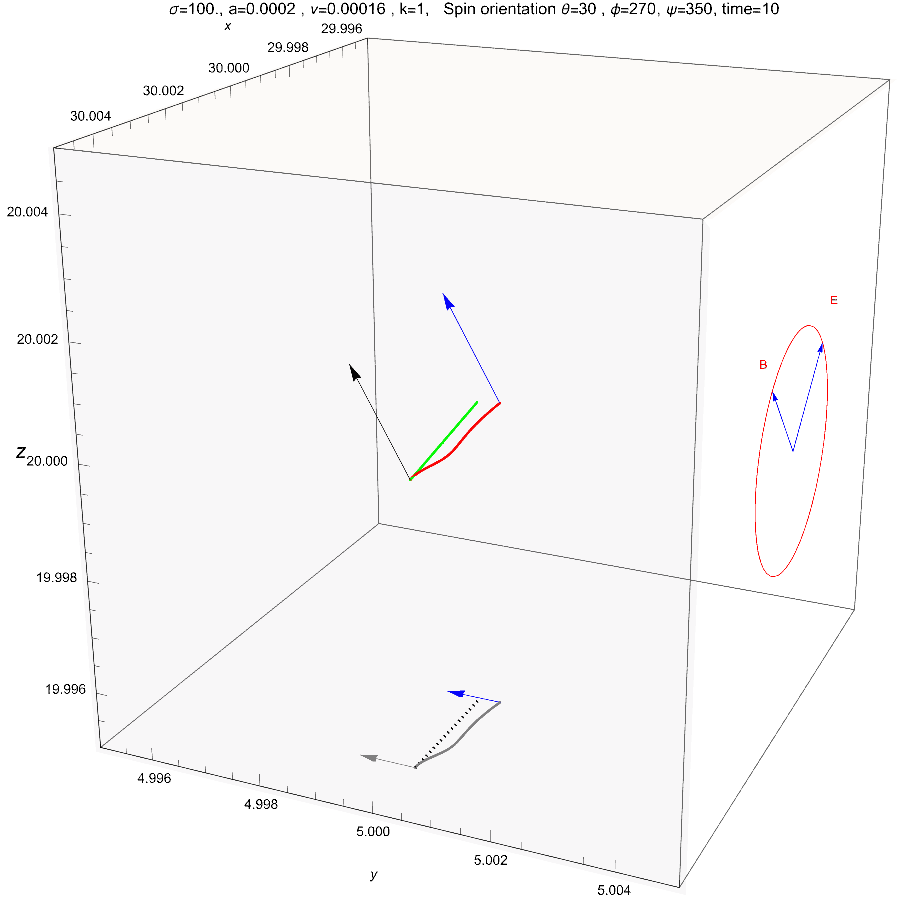}\\\includegraphics[width=5cm]{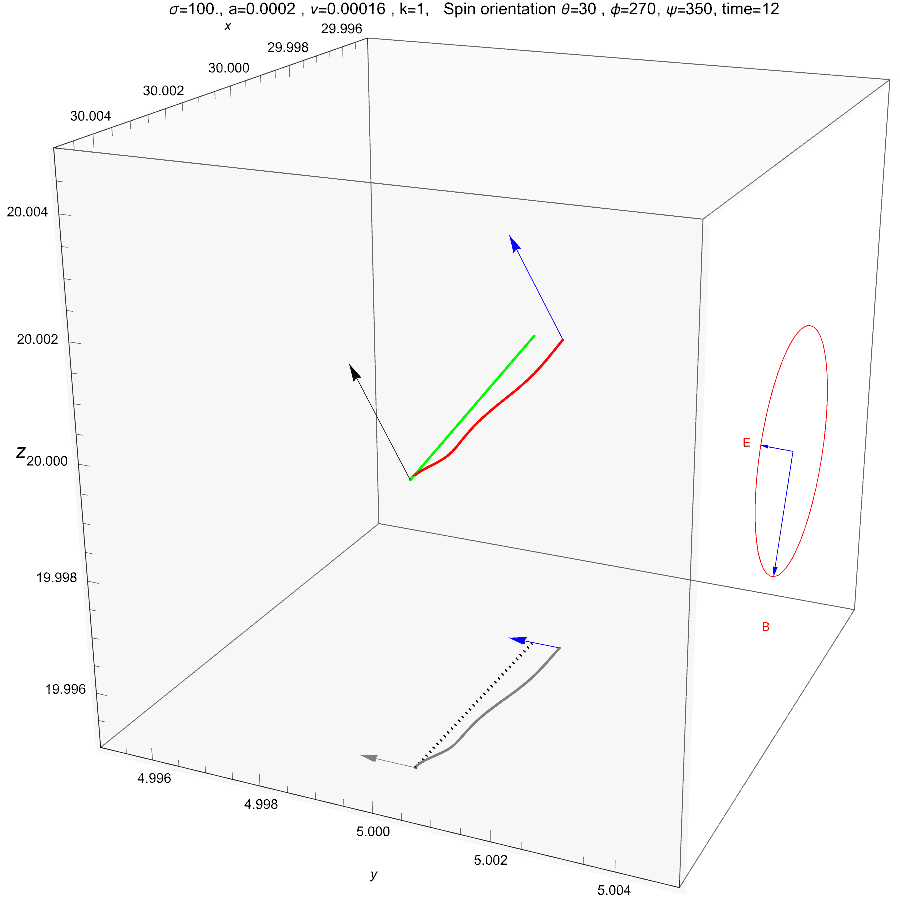}\includegraphics[width=5cm]{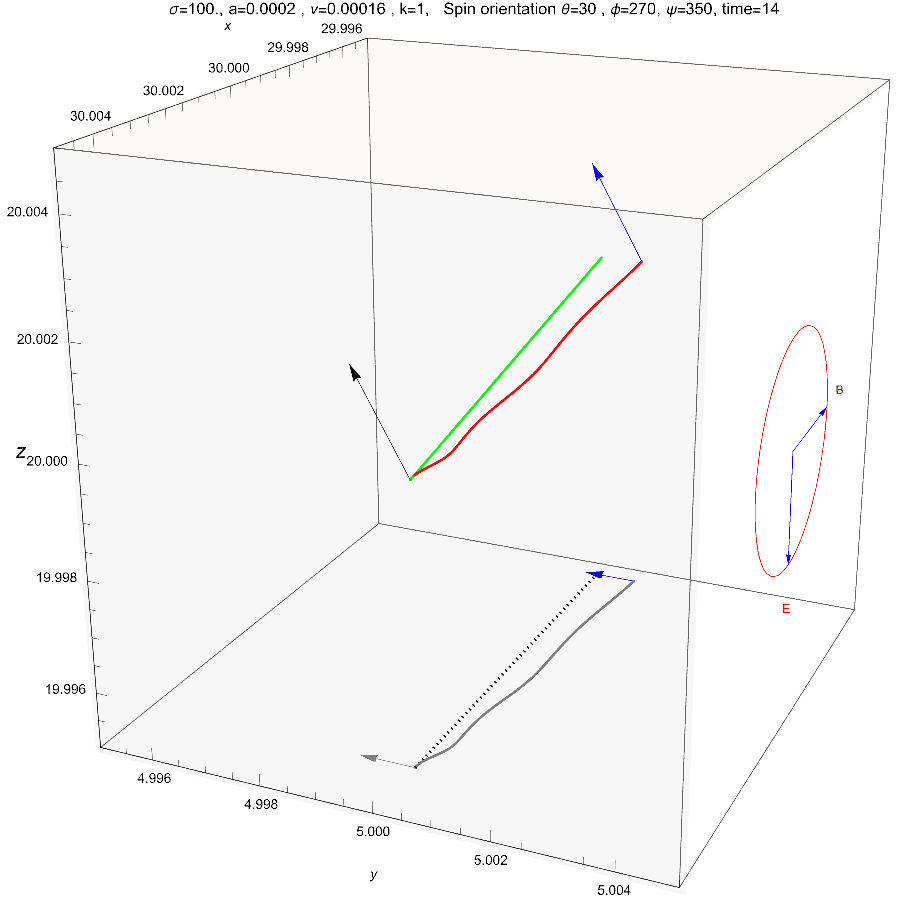}
\caption{Sequence of motions of the CM of the Dirac and point particle with $a=0.0002$, $\nu=0.00016$, $\sigma=100^\circ$ and the orientation of the spin is $\theta=30^\circ$, $\phi=270^\circ$ and the initial phase of the CC is $\psi=350^\circ$. The CM of the Dirac particle depicted in red is located initially at (30.0035,5,20) as well as the point particle in green. The front wave reaches the position of the CM of the particles at time $t=5$. The pictures represent the trajectories of both CM's at natural times $8, 10, 12$ and $14$, respectively. In the first picture the front wave has already arrived to the CM of both particles that move along different paths. The second, third and fourth pictures show that the CM motions of both particles are different. We have also depicted the initial (black) and final (blue) spins of the Dirac particle and the projection of the trajectories of the CM's of both particles and spins (grey for Dirac and dotted black for the point particle) at the bottom of the image. For these short times the point particle is moving on a plane orthogonal to the incoming direction of the wave and the direction depends on the value of $\sigma$, while the Dirac particle has a component of the linear momentum along the direction of the propagation of the wave. For longer times both particles move basically on an orthogonal plane to the motion of the wave with a small component of its linear momentum along the wave propagation direction. And this is independent of the electric charge of the particle.}  
\label{PPandDirac}
\end{figure}

\begin{figure}[!hbtp]\centering%
 \includegraphics[width=5cm]{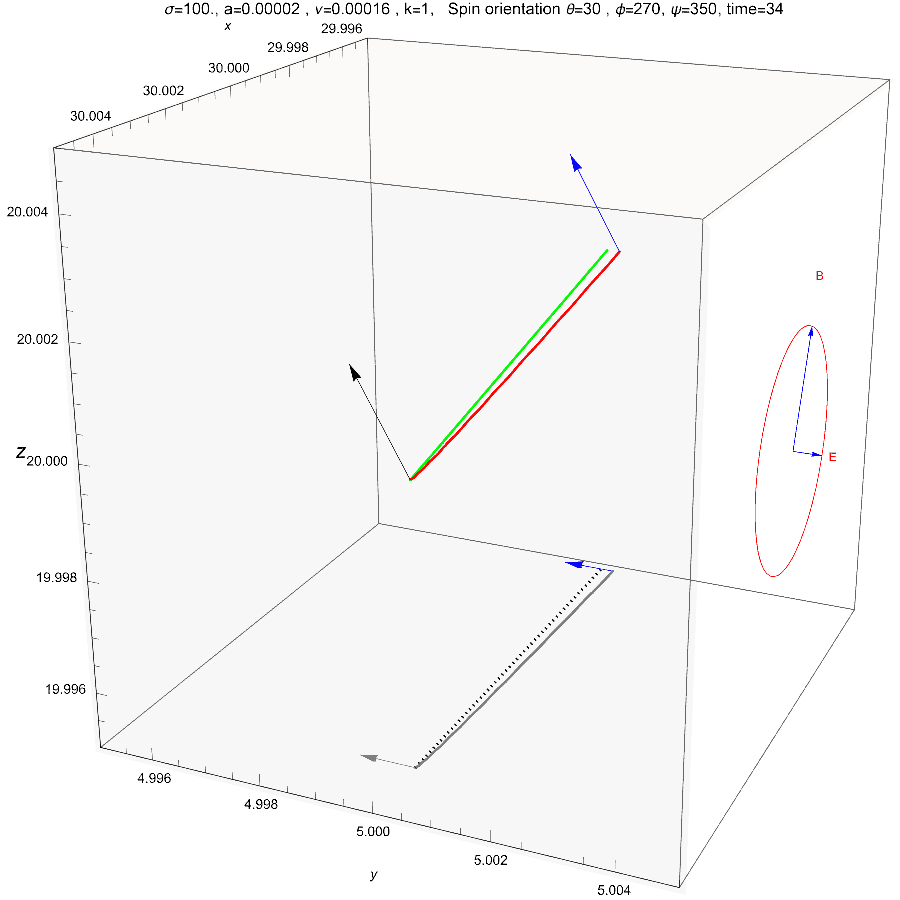}\includegraphics[width=5cm]{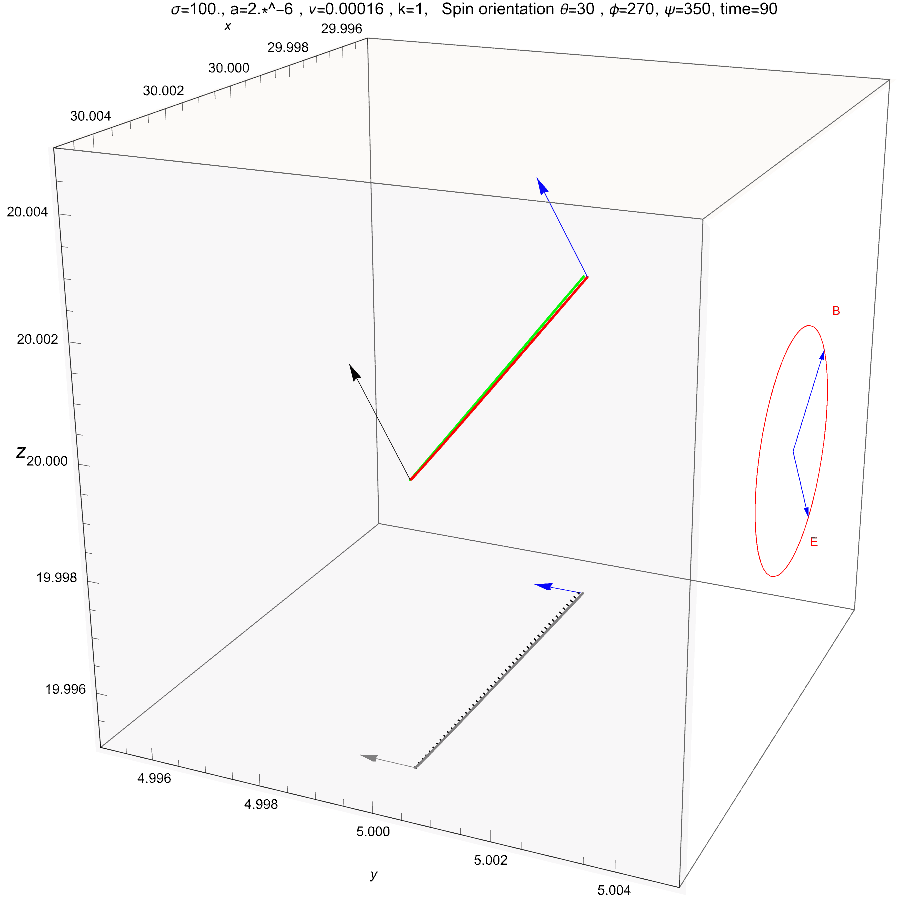}\includegraphics[width=5cm]{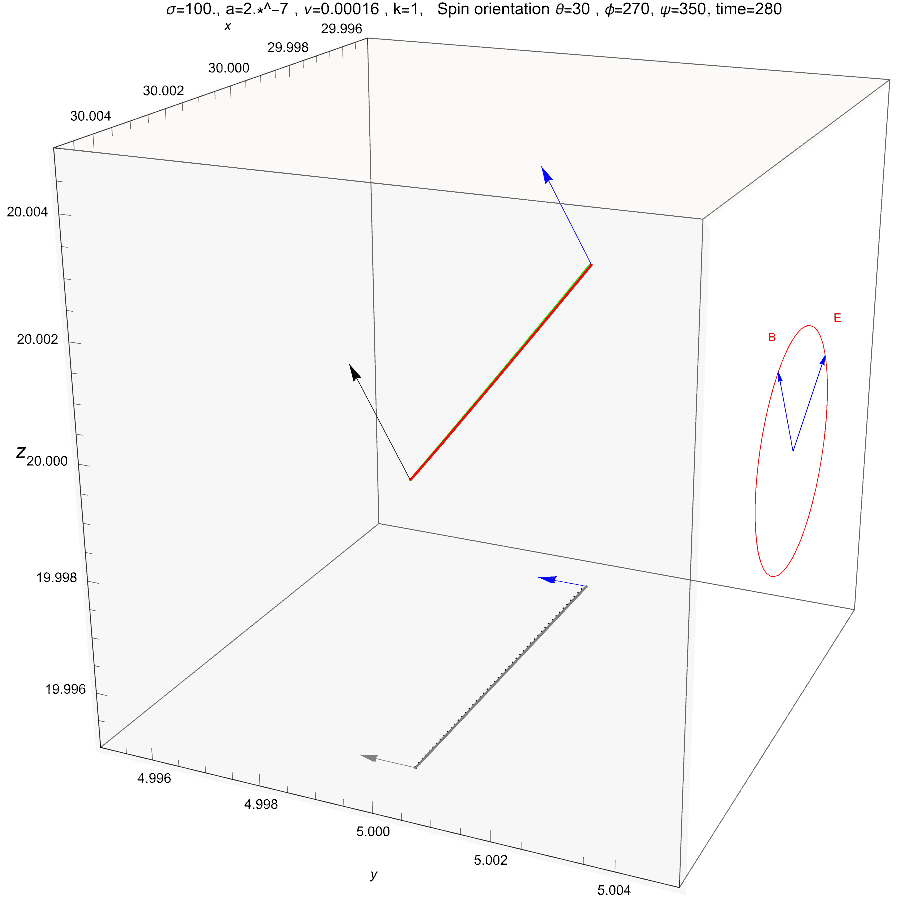}
\caption{Sequence of motions of the CM of the Dirac and point particle with the same boundary values as in the figure {\bf\ref{PPandDirac}} but decreasing in steps of ten the value of the interaction parameter $a$, $a=2\cdot10^{-5}$, $a=2\cdot10^{-6}$ and $a=2\cdot10^{-7}$, with greater integration times.}  
\label{PPandDirac3}
\end{figure}

\begin{figure}[!hbtp]\centering%
 \includegraphics[width=5cm]{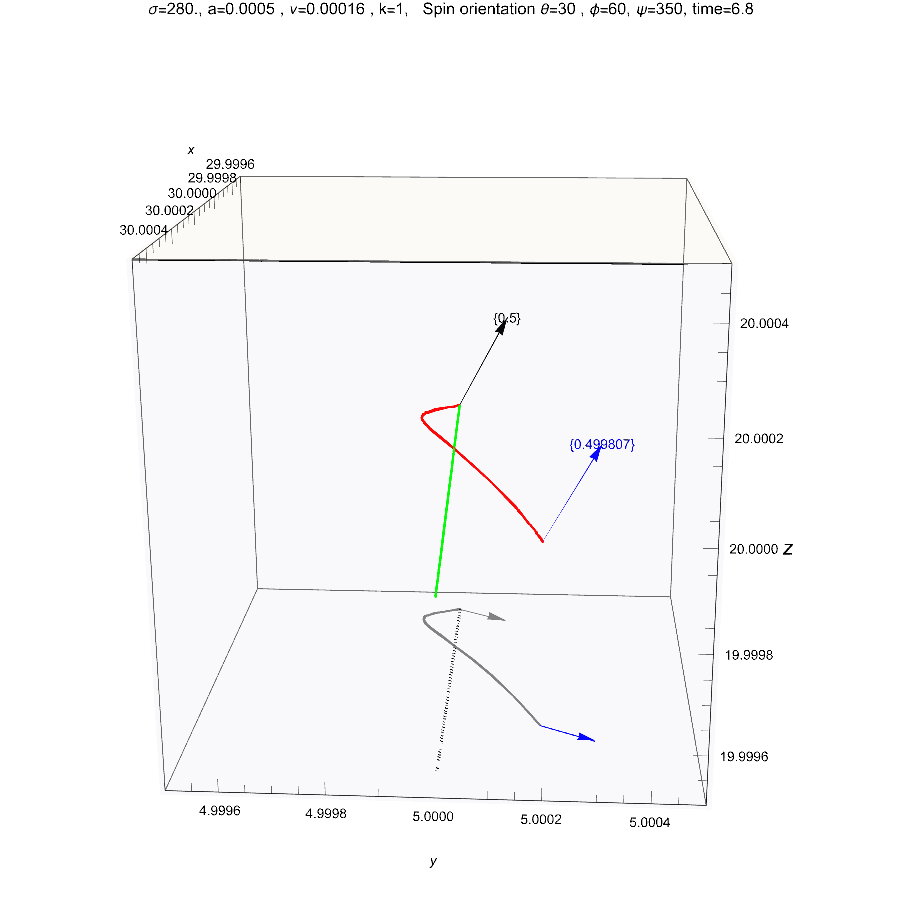}\includegraphics[width=5cm]{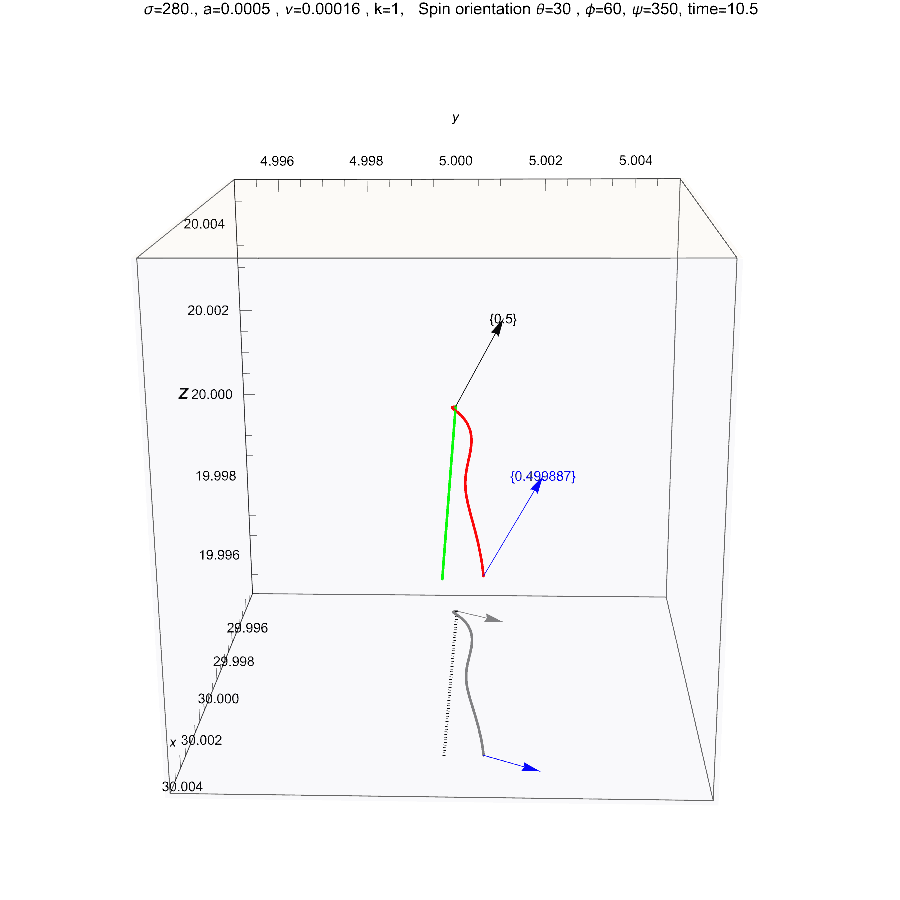}\includegraphics[width=5cm]{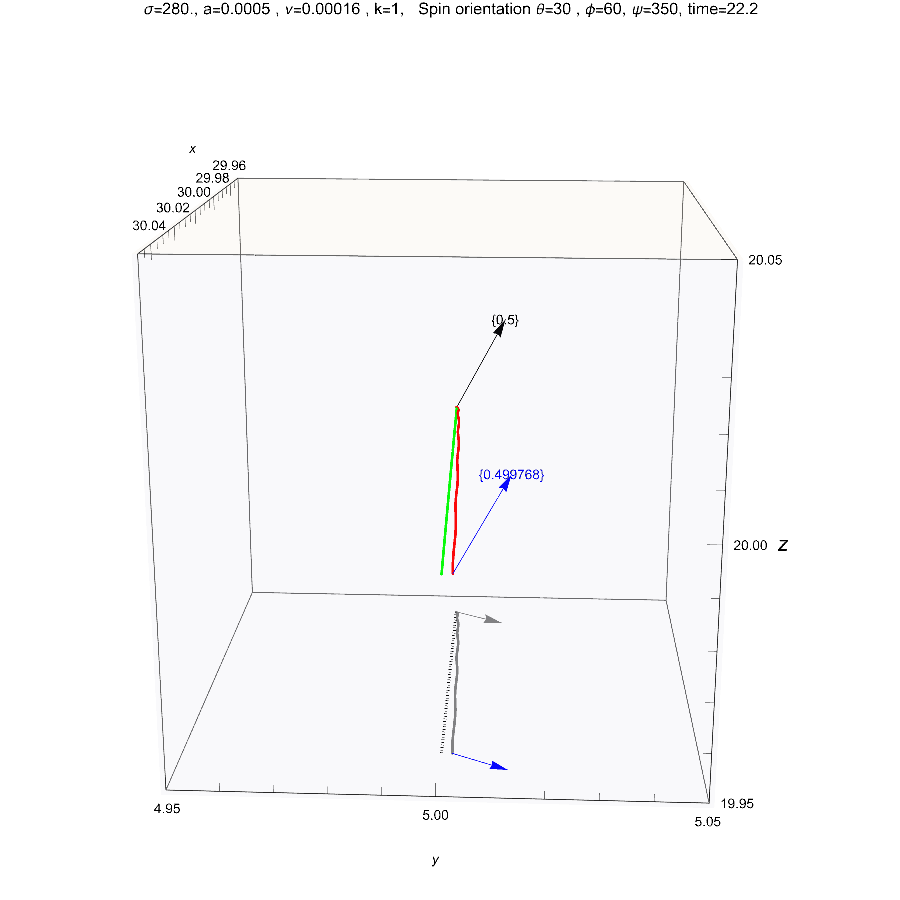}
\caption{Sequence of motions of the CM of the Dirac and point particle with $a=0.0005$, $\nu=0.00016$, $\sigma=280^\circ$, $180^\circ$ more than in the figure {\bf\ref{PPandDirac}}, the CM motions are in the opposite direction. The initial phase of the CC is the same $\psi=350^\circ$. The CM of the Dirac particle depicted in red is located initially at (30,5,20) as well as the point particle in green. The pictures represent the trajectories of both CM's at natural times $6.8, 10.5$ and $22.2$, respectively. We have enlarged the scale ten times in the second picture and hundred times in the third to see the progress of the CM's. We have also depicted the initial (black) and final (blue) spins of the Dirac particle and the projection of the trajectories of the CM's of both particles (grey for Dirac and dotted black for the point particle) at the bottom of the image. For these short times the point particle is moving on a plane orthogonal to the incoming direction of the wave and the direction depends on the value of $\sigma$. In the first picture the CM of the Dirac particle goes backward but in few time units gets a component of its linear momentum along the wave trajectory. This linear momentum transfer is greater than for the point particle.}  
\label{PPandDirac2}
\end{figure}

In the figure {\bf\ref{PPandDirac}} we show the early instants of the evolution when the front has already arrived to the position of the two particles. The spatial scale have been enlarged to appreciate the differences.
\newpage
With the same boundary variables and parameters of the figure {\bf\ref{PPandDirac}} we depict in the figure {\bf\ref{PPandDirac3}} the integration by changing the interaction parameter $a=2\cdot10^{-5}, 2\cdot10^{-6}$ and $2\cdot10^{-7}$, respectively and longer integration times.
\newpage
With the same boundary conditions as in the figure {\bf\ref{PPandDirac}} we enlarge in $180^\circ$ the initial phase of the field $\sigma$ and obtain the  trajectories of the figure {\bf\ref{PPandDirac2}}, where the orientation of the trajectories have changed.

\newpage

\section{Conclusions}
\label{Conclusions}
The interaction of a circularly polarized electromagnetic plane wave with a classical Dirac particle at rest has been analyzed with a classical model of a spinning particle that when quantizes satisfies Dirac's equation. From the classical point of view represents a classical way of describing the interaction of an electromagnetic plane wave with an electron. It has been impossible to obtain analytical solutions of the non-linear dynamical equations and therefore the analysis is done on the numerical solutions of the system of ordinary differential equations which define the trajectories of the CM and CC of the Dirac particle. The equations have been written in natural units, the different variables are dimensionless, and the intrinsic parameters of the electron are: mass $m=1$, spin $S=1/2$, electric charge $e=\pm1$, magnetic dipole moment
$\mu_B=1/2$ and electric dipole moment $d=1/2$. The universal constants take the values $c=1$ and $\hbar=1$.
 
As analyzed in Section {\bf\ref{Variation}}, because the Dirac particle is initially at rest, we have to wait a long integration time to obtain a relevant increase of the CM velocity to appreciate a modification of the CM spin of the particle. 
The analyzed numerical experiments are determined for very high intensity of the interaction parameter $a$ to appreciate at the length and time scale of the Dirac particle the evolution of both points the CC and the CM.
The numerical program allows us to determine the variation of the absolute value of the CM spin during the integration time.

For the direction of the motion of the CM, the internal phase $\psi$ of the CC has no influence except in the preliminary instants of the interaction, while it is the internal phase of the wave $\sigma$ that determines the global direction of this motion. Irrespective of the initial spin orientation of the Dirac particle the motion of the CM is basically in the same direction.

When compared with the point particle, the Dirac particle has a different behavior when interacting with the plane wave at the early stages of the interaction and in the long term the evolution is slightly different. The plane wave transfers to the Dirac particle energy, linear momentum and angular momentum. From the physical point of view  a circularly polarized electromagnetic plane wave represents a mechanical system of a huge number of monochromatic photons all of them with the same spin orientation, forward or backward with respect to the direction of propagation of the wave, as was shown experimentally by Beth in 1936 \cite{Beth}. A very glancing interaction appears to take place, where the Dirac particles are scattered in a direction transverse to the direction of the wave propagation. In general, all motions have a small positive linear momentum transfer along the direction of the wave propagation, irrespective of the sign of the charge of the particle. 

If we compare this result with the classical Compton effect of the interaction of a photon and a pointlike electron it seems that the scattering angle of the electron is $\beta\approx 90^\circ$. The outgoing plane wave remains unchanged after the interaction as if the presence of the Dirac particle did not modify the electromagnetic field of the wave.  Nevertheless, the classical Compton effect analyzes the scattering of a single point electron with a single photon. In the plane wave interaction when the wave arrives to the location of the CC of the Dirac particle we have a continuous interaction of the electromagnetic field of the wave packet with the particle, thus increasing the energy and linear momentum with no limit. This kind of interaction seems to be different than the interaction of a beam of photons with an electron at rest, even if we admit multiple scattering of the electron with consecutive photons.

The classical analysis of the Compton effect of the interaction of a spinning photon with a spinning Dirac particle is left to a future paper. Electromagnetism is invariant under translations and rotations and their interactions conserve energy, linear momentum and angular momentum. In addition to energy and linear momentum conservation of the usual analysis of the classical Compton effect we shall consider also the angular momentum conservation, by including the possible variation of the spins and orbital angular momenta of both particles.
 

\end{document}